\documentclass[journal,10pt,twocolumn,twoside]{IEEEtran}
\usepackage{mathtools}
\usepackage{empheq}
\usepackage{algpseudocode, algorithm}
\usepackage{algorithmicx}
\usepackage{subfig}
\usepackage{url}
\usepackage{amssymb}
\usepackage{amsthm}
\usepackage{amsmath}
\usepackage{changes}
\usepackage{cite}
\usepackage{multirow}
\usepackage{bm}
\usepackage{hyperref}
\usepackage{balance}
\usepackage{xcolor}
\usepackage{makecell}
\usepackage{amsbsy}
\usepackage{float}
\usepackage{adjustbox}
\usepackage{forest}

\newcolumntype{P}[1]{>{\centering\arraybackslash}p{#1}}

\DeclareMathOperator*{\argmin}{argmin}

\ifCLASSINFOpdf
\else
\fi
\begin{document}
%
\title{Swarm Intelligence for Next-Generation Wireless Networks: Recent Advances and Applications}

\author{Quoc-Viet Pham, Dinh C. Nguyen, 
Seyedali Mirjalili, \\Dinh Thai Hoang, Diep N. Nguyen, Pubudu N. Pathirana, and Won-Joo Hwang
\thanks{Quoc-Viet Pham is with Research Insitute of Computer, Information and Communication, Pusan National University, 2, Busandaehak-ro 63beon-gil, Geumjeong-gu, Busan 46241, Korea (e-mail: vietpq@pusan.ac.kr).}

\thanks{Dinh C. Nguyen and Pubudu N. Pathirana are with School of Engineering, Deakin University, Waurn Ponds, VIC 3216, Australia (e-mails: {cdnguyen, pubudu.pathirana}@deakin.edu.au).} 

\thanks{Seyedali Mirjalili is with Center for Artificial Intelligence Research and Optimization, Torrens University Australia, 90 Bowen Terrace, Fortitude Valley, Brisbane, QLD 4006, Australia (e-mail: seyedali.mirjalili@griffithuni.edu.au).}

\thanks{Dinh Thai Hoang and Diep N. Nguyen are with School of Electrical and Data Engineering, University of Technology Sydney, Ultimo, NSW 2007, Australia (e-mail: \{diep.nguyen, hoang.dinh\}@uts.edu.au).}

\thanks{Won-Joo Hwang is with School of Biomedical Convergence Engineering, Pusan National University, 49, Busandaehak-ro, Mulgeum-eup, Yangsan-si, Gyeongsangnam-do 50612, Korea (e-mail: wjhwang@pusan.ac.kr).}

}

\markboth{Swarm Intelligence for Next-Generation Wireless Networks: Recent Advances and Applications}{Swarm Intelligence for Next-Generation Wireless Networks: Recent Advances and Applications}


%


\IEEEtitleabstractindextext{%
\begin{abstract}
Due to the proliferation of smart devices and emerging applications, many next-generation technologies have been paid for the development of wireless networks. Even though commercial 5G has just been widely deployed in some countries, there have been initial efforts from academia and industrial communities for 6G systems. In such a network, a very large number of devices and applications are emerged, along with heterogeneity of technologies, architectures, mobile data, etc., and optimizing such a network is of utmost importance. 
Besides convex optimization and game theory, swarm intelligence (SI) has recently appeared as a promising optimization tool for wireless networks. As a new subdivision of artificial intelligence, SI is inspired by the collective behaviors of societies of biological species.
In SI, simple agents with limited capabilities would achieve intelligent strategies for high-dimensional and challenging problems, so it has recently found many applications in next-generation wireless networks (NGN). 
However, researchers may not be completely aware of the full potential of SI techniques. 
In this work, our primary focus will be the integration of these two 
domains: NGN and SI. 
Firstly, we provide an overview of SI techniques from fundamental concepts to well-known optimizers. 
Secondly, we review the applications of SI to settle emerging issues in NGN, including spectrum management and resource allocation, wireless caching and edge computing, network security, and several other miscellaneous issues. 
Finally, we highlight open challenges and issues in the literature,
and introduce some interesting directions for future research.
\end{abstract}

\begin{IEEEkeywords}
5G and Beyond, 6G, Artificial Intelligence (AI), Computational Intelligence, Swarm intelligence (SI), Next-Generation Wireless Networks.
\end{IEEEkeywords}}

%
\maketitle
\IEEEdisplaynontitleabstractindextext
\IEEEpeerreviewmaketitle

\section{Introduction}
\label{Sec:Introduction}
Four generations of cellular networks have been introduced since the inception of the first generation in early 1980s. 
\textcolor{black}{It is expected that massive numbers of connected devices, applications, technologies, network architecture, etc. will be available in next-generation wireless networks (NGN, i.e., fifth-generation (5G) and beyond) \cite{Pham2020Survey_MEC}.} At the time we complete this survey (2020 August), there have been some 5G commercial deployments based on the New Radio (NR) standard \cite{Ericsson2020_commercial5G}.
The pace of deployment would be accelerated by the completion of 3rd Generation Partnership Project (3GPP) Release 16 specification with more details and enhancements concerning industrial Internet of Things (IoT), satellite integration, NR-based unlicensed spectrum, enhanced multi-input multi-output (eMIMO), advanced vehicle to everything (V2X) support, ultra-reliable and low latency communication (uRLLC) enhancements, and security \cite{3GPP_TR21_916}. In summary, three main use cases supported in 5G wireless systems are enhanced mobile broadband (eMBB), massive IoT, and uRLLC. 

Besides many benefits offered by NGN (e.g., higher data rates and reliability, lower latency, and more capacity), there are many challenges needed to be tackled to make it successful (e.g., the network operators need to reduce their expenditure and complexity of operations, and how to support more use cases and services from the third-party companies). \textcolor{black}{There are many problems needed to be studied in NGN such as joint power control and user clustering in non-orthogonal multiple access (NOMA) networks \cite{Dai2018Survey_NOMA}, task offloading and resource allocation in multi-access edge computing (MEC) systems \cite{Mao2017_aSurveyMEC}, pilot allocation in V2X vehicular communications and MIMO \cite{Trinh2018JointPilot}, network slicing in virtualized networks \cite{Tun2019WirelessNetwork}, and beamforming in (cell-free massive) MIMO systems \cite{Pham2019Revisiting}. Several powerful techniques, e.g., convex optimization, game theory, and machine learning have been used to solve these problems. As a suitable alternative, swarm intelligence (SI), a subset of artificial intelligence (AI), has been recently and widely used in the literature with impressive performance. Despite promising results, we are not aware of any survey dedicated to the use of SI techniques for emerging issues in NGN. Filling this gap, \textit{we carry out a contemporary survey on the applications of swarm intelligence (SI) techniques for NGN}. The investigation on emerging issues includes spectrum management and resource allocation, wireless caching and edge computing, network security, and miscellaneous issues.}

\subsection{\textcolor{black}{A Brief On Optimization Tools}}
\textcolor{black}{Various methods have been applied to address emerging issues in NGN. We briefly describe these techniques as follows.}
\begin{itemize}
	\item \textit{Convex Optimization}: In the event that an optimization problem is convex, some well-know methods and off-the-shelf solvers, for example, interior-point method and CVX \cite{Grant2014CVX} would be used to get the optimal solution. The convexity feature is rarely available and scholars typically utilize some approaches to convexify the underlying problem or approximate it as a sequence of convex problems, e.g., difference of two convex functions (DC) programming \cite{Kha2012FastGlobal},  semidefinite relaxation (SDR) \cite{Luo2010SemidefiniteRelaxation}, and second-order cone programming (SOCP) \cite{Tran2012FastConverging}. Nevertheless, typically the computational complexity increases at an exponential rate with the optimization dimensionality, e.g., the numbers of IoT devices and the quantization levels of channel states. In such a case, heuristic algorithms are highly preferred thanks to their features of simplicity and low complexity. For example, the heuristic method can relax a binary variable as a continuous one, and similarly a binary constraint as a quadratic one. However, the main drawback of such  heuristic schemes is that the convergence to optima and the existence of the optimality are not guaranteed.
	 
	\item \textit{Game Theory}: The purpose of game theory is to study the interactions among independent and rational agents, e.g., mobile users in a cellular network and edge hosts in MEC systems. Game theory has a variety of applications in various disciplines such as economics, business, philosophy, and recently in wireless networks. The design of efficient algorithms for large-scale, distributed, dynamic, and heterogeneous wireless systems, is considered as the primary use of game theory. For instance, coalitional game theory is used for joint optimization of the network throughput and content service satisfaction degree in cashing systems \cite{Xu2019NetworkCoding}, channel allocation in device-to-device (D2D) communications \cite{Wang2019SubchannelAllocation}, and user pairing in NOMA systems \cite{Wang2019UserClustering}. Recently, transport and matching theories, as powerful mathematical frameworks, have been used for many optimization problems (e.g., cell association, computation offloading decision, and interference management) \cite{Mozaffari2017Wireless, Pham2018Decentralized}. In spite of clear advantages, game theory is usually reliant on the assumption of rationality and the underlying models. These assumptions can affect the game strategies and final outcomes, so game theory is not a suitable tool for some network settings and scenarios.
	
	\item \textit{Machine Learning Approaches}: Over recent decades, machine learning (ML) plays a key role for many revolutionary changes in many areas, e.g., computer vision, signal processing, agriculture, bioinformatics, and cheminformatics. More recently, ML has been emerging in mobile communications and networking arena, and there have been some surveys and tutorials on ML \cite{sun2019application, zhang2019deep, lim2020federated}. Additionally, to promote the use of ML for the physical and medium access control layers in wireless and communication networks, IEEE communications society (IEEE ComSoc) has started an initiative called ``Machine Learning for Communications Emerging Technology Initiative (ETI)" (\url{https://mlc.committees.comsoc.org/}), where news, datasets, competitions, research library, simulation codes, etc. are available and continuously updated. However, ML is not always applicable to wireless and communication networks due to several challenges. For example, creating high-quality and standard datasets for a new network architecture would be a nontrivial task and may not be feasible for many cases. Further, training deep neural networks with a very large dataset and massive optimization parameters is time-consuming and usually transferred to off-line training \cite{zhang2019energy}.
\end{itemize}
Other tools for modeling and analyzing wireless networks can be stochastic geometry and random graphs, which have been found in problems such as interference characterization and outage probability, throughput, and packet error rate. However, from an optimization perspective, these tools are not covered in this subsection. Interested readers are suggested to read the tutorial \cite{Haenggi2009Stochastic} and the 
book \cite{Baccelli2010Stochastic_II}.

\subsection{\textcolor{black}{Swarm Intelligence for Next-Generation Wireless Networks}}
The main challenge in solving 
NP-hard 
problems is an exponential increase in the complexity levels (e.g., in time and memory). In such cases, algorithms that find exact solutions (i.e., exact algorithms) are not efficient. Especially in NGN, the massive numbers of connected devices, base stations (BSs), channel states, heterogeneous resources, etc. increase the challenges of finding the exact algorithms. Simple examples of such problems are mobility-aware user association in ultra-dense networks, signal detection and channel estimation, and routing in IoT networks. 

\begin{figure*} [t]
	\centering
	\includegraphics[width=0.80\textwidth]{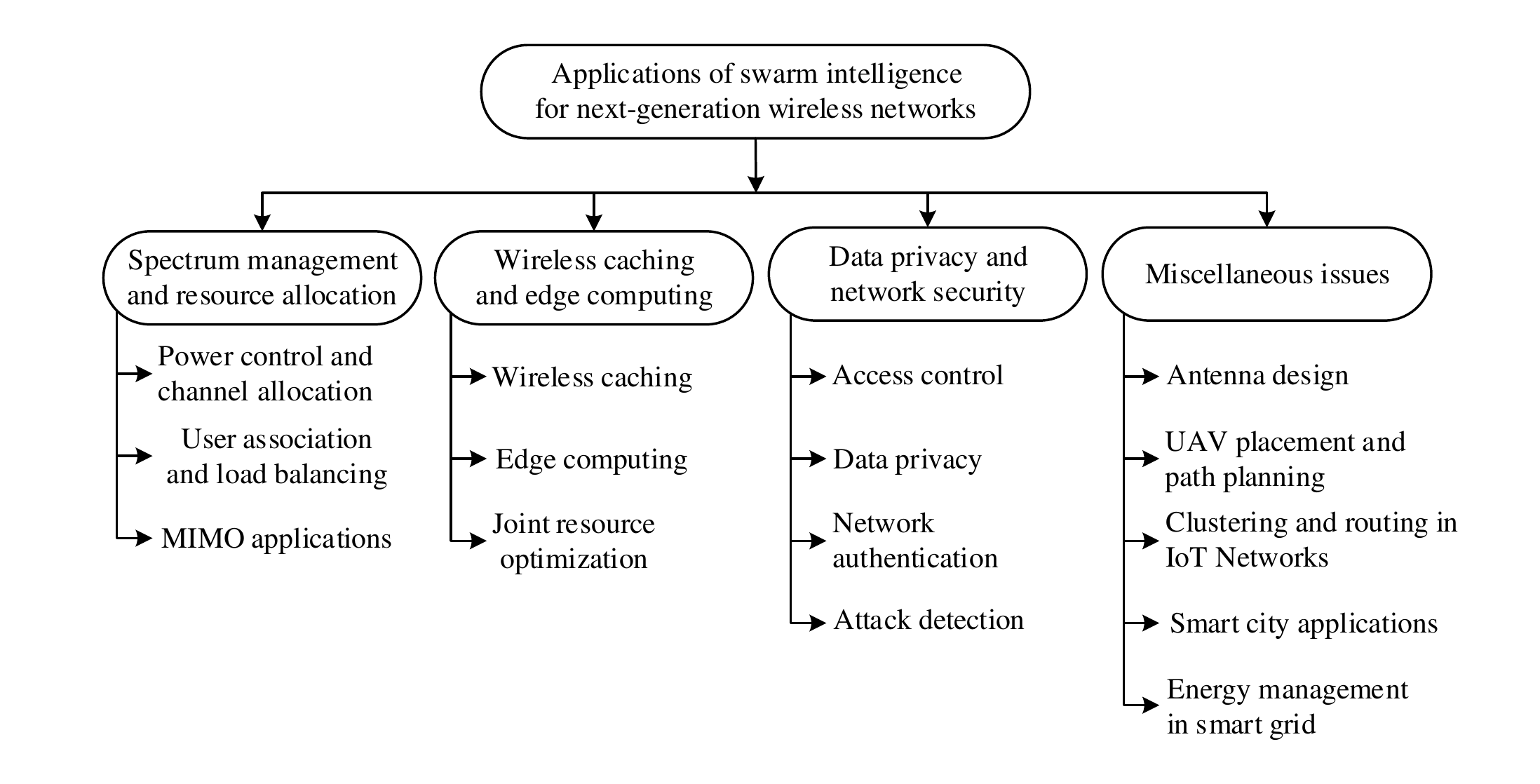}
	\caption{A classification of the applications of swarm intelligence for next-generation wireless networks.}
	\label{Fig:Taxonomy_of_Applications}
\end{figure*}

\textcolor{black}{
As a subset of bio-inspired and nature-inspired algorithms,
SI studies complex collective behavior of the systems composed of many simple components, which can interact with other agents locally and with their surrounding environment \cite{martens2011editorial}. For example, the particle swarm optimization (PSO) is designed to mimic the movement of organisms in flocks of birds, and the ant colony optimization (ACO) is inspired by the behavior of ants in finding the shortest path to the food sources. The main advantages of SI and other nature-inspired algorithms over conventional approaches like gradient-based and game-theoretic algorithms are as follows:
\begin{enumerate}
    \item \textit{no assumptions about the problem being optimized},
    \item \textit{ability to find high-quality solutions by balancing between exploration and exploitation},
    \item \textit{no need for the gradient information of the problem being optimized (i.e., gradient-free)},
    \item \textit{simplicity and easy implementation}.
\end{enumerate}
These particularities have made SI to be a promising solution for many problems in various research areas, and therefore have been applied to solve 
an increasing number of problems in NGN. For example, an ACO algorithm is proposed to optimize the mobility trajectory for a mobile sink in home automation networks \cite{Wang2015Bioinspired}, the PSO method is employed to optimize antenna placement in distributed MIMO systems in \cite{Forooshani2014Optimization}, and the Firefly algorithm (FA) is utilized to find the key update and residence management for an enhanced handover security \cite{Vien2019EnhancingSecurity}. The work in \cite{pham2020sum} employed the Harris hawks optimizer (HHO) to overcome the non-convex difficulties caused by the channel model in unmanned aerial vehicle (UAV) assisted visible light communications. The HHO-based algorithm proposed in \cite{pham2020sum} can solve the optimization problem by finding multiple variables simultaneously and is not specific to any convex approximation technique as typically used in existing studies. Another application of SI can be found in \cite{Jiang2020Deep}, where the PSO is adopted to solve a resource scheduling problem in UAV-MEC systems and generate training samples for a deep neural network. The SI mechanisms used in the literature above have shown many promising results, e.g., remarkable performance with high reliability and convergence guarantee. This observation stimulates the increasing use of SI as a vital tool for solving emerging issues in NGN. 
}

\subsection{\textcolor{black}{State-of-the-art Surveys and Our Contributions}}
Although there are some surveys related to SI,  their focuses are not on the applications of SI in NGN. Specifically, there have been existing surveys, such as \cite{martens2011editorial, Kolias2011SwarmIntelligence, Mavrovouniotis20171SurveyonSI, Bai2006SurveySI,Tan2016SurveyGPU, Ertenlice2018SurveySI}, but they are specifically for other topics rather than NGN, e.g., data mining \cite{martens2011editorial}, intrusion detection \cite{Kolias2011SwarmIntelligence}, dynamic optimization \cite{Mavrovouniotis20171SurveyonSI}, electric power system \cite{Bai2006SurveySI}, graphical processing unit (GPU)-based implementation \cite{Tan2016SurveyGPU}, and portfolio optimization \cite{Ertenlice2018SurveySI}. There have been also surveys discussing the applications of SI for wireless networks such as \cite{Saleem2011Swarm, Zhang2014OnSwarm, balusamy2015bio, Primeau2018Review_CI, Anandakumar2018BioInspired}. However, they do not focus on next-generation wireless systems. In particular, 
the survey in \cite{Saleem2011Swarm} is
dedicated to the routing mechanisms in wireless sensor networks (WSNs), the survey in \cite{Zhang2014OnSwarm} presents SI for self-organized networking, the survey in \cite{balusamy2015bio} presents bio-inspired mechanisms in cloud computing like resource scheduling, load balancing, and security, the surveys in \cite{Primeau2018Review_CI} provide reviews of various problems (e.g., data aggregation, sensor fusion, scheduling, and security) in WSNs using computational intelligence approaches, and the survey in \cite{Anandakumar2018BioInspired} reviews bio-inspired SI for social cognitive radio network applications. Moreover, there have been some surveys dedicated to the applications of a specific SI technique, for example, \cite{Zhang2017Survey_ACO} provides surveys on theoretical results on the ACO and its applications to routing protocols in WSNs. \textcolor{black}{The fundamentals of PSO and its application to WSNs can be found in \cite{Zhang2015aComprehensiveSurvey}. Recent years have seen various surveys on other AI techniques for NGN such as machine learning for wireless networks \cite{sun2019application}, deep learning for mobile networking \cite{zhang2019deep}, and federated learning for mobile edge networks \cite{lim2020federated}. The surveys on artificial bee colony (ABC), firefly algorithm (FA), grey wolf optimizer (GWO), whale optimization algorithm (WOA), and HHO are presented in \cite{karaboga2014comprehensive, Fister2013survey_FA, faris2018grey, Gharehchopogh2019aSurvey_WOA, heidari2019harris}.}

In summary, the surveys mentioned above either consider the applications of SI for other topics,
discuss the applications of SI for conventional problems in wireless networks (e.g., routing protocols in WSNs and self-organized networking), and review other AI disciplines (e.g., machine learning, deep learning, and federated learning). We are not aware of any survey that specifically discusses the applications of SI for emerging issues in NGN. Moreover, almost all these surveys were published many years ago, so the topics discussed are not up-to-date. These observations motivate us to conduct this survey article with the fundamentals of SI and an up-to-date review of the applications of SI to address emerging issues in NGN. For convenience, three major issues in NGN are reviewed in this survey, including spectrum management and resource allocation, wireless caching and edge computing, network security, as shown in Fig.~\ref{Fig:Taxonomy_of_Applications}. 
Also, a percentage of literature related to a number of miscellaneous yet important issues is reviewed: antenna design, UAV placement and path planning, clustering and routing in IoT networks, smart city applications, and energy management in smart grid.

\begin{table*}[t]
	\centering
	\caption{List of the acronyms.}
	\label{table:TableAcr}
	{\renewcommand{\arraystretch}{1.4}
	\resizebox{\linewidth}{!}{
		\begin{tabular}{ll|ll|ll}
			\hline
			\textbf{Acronyms} & \textbf{Meaning} & \textbf{Acronyms} & \textbf{Meaning} & \textbf{Acronyms} & \textbf{Meaning} \\ \hline
			5G & Fifth-Generation &
			6G & Sixth-Generation &
			ABC & Artificial Bee Colony \\ \hline 
			ACO & Ant Colony Optimization &
			AI & Artificial intelligence &
			BS & Base Station \\ \hline
			ACO & Ant Colony Optimization &
			D2D & Device-to-Device &
			FA & Firefly Algorithm \\ \hline
			GA & Genetic Algorithm &
			GWO & Grey Wolf Optimizer &
			HetNet & Heterogeneous Network \\ \hline
			HHO & Harris Hawks Optimizer &
			IoT & Internet of Things &
			LTE & Long Term Evolution \\ \hline
			MANET & Mobile Ad Hoc Network &
			MEC & Multi-Access Edge Computing &
			MIMO & Multi-Input Multi-Output \\ \hline
			MLD & Maximum Likelihood Detection &
			mmWave & millimeter-Wave &
			NGN & Next-Generation Wireless Networks \\ \hline
			NOMA & Non-Orthogonal Multiple Access &
			OMA & Orthogonal Multiple Access &
			PSO & Particle Swarm Optimization \\ \hline 
			QoS & Quality of Service &
			QAM & Quadrature Amplitude Modulation &
			RF & Radio Frequency  \\ \hline 
			SDN & Software-Defined Networking &
			SER & Symbol Error Rate &
			SI & Swarm Intelligence \\ \hline 
			UAV & Unmanned Aerial Vehicle &
			WOA & Whale Optimization Algorithm &
			WSN & Wireless Sensor Network \\
			\hline 
		\end{tabular}
	}
	}
\end{table*}

In summary, the contributions of this survey paper are described as follows.
\begin{itemize}
	\item Firstly, we present fundamental knowledge of SI and provide a review of representative techniques, including most well-known SI approaches: PSO, ABC, ACO, FA, and recent SI advances: GWO, WOA, and HHO. 
	
	\item Secondly, we discuss the applications of SI techniques for major issues in NGN, including clustering and routing in IoT systems, spectrum management and resource allocation, wireless caching and edge computing, network security. 
	
	\item Finally, we highlight a number of important challenges and future research directions for the use of SI in NGN. 
\end{itemize}

\subsection{\textcolor{black}{Paper Organization}}
The rest of paper is organized as follows. Section~\ref{Sec:SI_Techniques} presents the fundamentals of SI techniques. Sections~\ref{Sec:SpectrumManagement}-\ref{Sec:DataPrivacy} provide a comprehensive survey on the applications of SI for emerging issues in NGN, spectrum management and resource allocation, wireless caching and edge computing, and network security. After that, miscellaneous issues using SI, such as antenna design, UAV placement and path planning, clustering and routing in IoT networks, smart city applications, and energy management in smart grid, are presented in Section~\ref{Sec:Miscellaneous_Issues}. Section~\ref{Sec:Challenges} outlines the challenges, open issues, and some future research directions. Finally, Section~\ref{Sec:Conclusion} concludes the survey paper. 
The list of frequently used acronyms is shown in Table~\ref{table:TableAcr}.
\section{Swarm Intelligence Techniques: An Overview}
\label{Sec:SI_Techniques}
In this section, we first present the fundamentals of evolutionary algorithms and SI techniques. Secondly, we show the general principles that SI deals with constrained and mixed-integer optimization problems. Finally, the summaries of most well-known SI approaches like PSO, ABC, ACO, and FA as well as recent advancements such as GWO, WOA, and HHO are presented. 

\subsection{\textcolor{black}{Fundamentals Of Swarm Intelligence And Stochastic Optimization}}
The term ``swarm intelligence" is first appeared in \cite{beni1993swarm} when cellular robotic systems are empowered by the intelligence of swarms. It refers to biological systems, in which multiple individuals achieve certain goals with local interaction without any centralized control unit. For instance, an ant colony is made of a queen and many workers, soldiers, babysitters, etc. The queen is just responsible for laying egg, but other ants construct the nest, raise larvae, defend the colony, collect food, etc. The queen is not a commander, and somehow tasks are coordinated and goals are achieved. This is an example of evident functionality pertaining to SI. Ants are proved to communicate locally using pheromone and solve complex problems to ensure the survival of their colony. 

\begin{figure*}[t]
	\centering
	\includegraphics[width=0.60\linewidth]{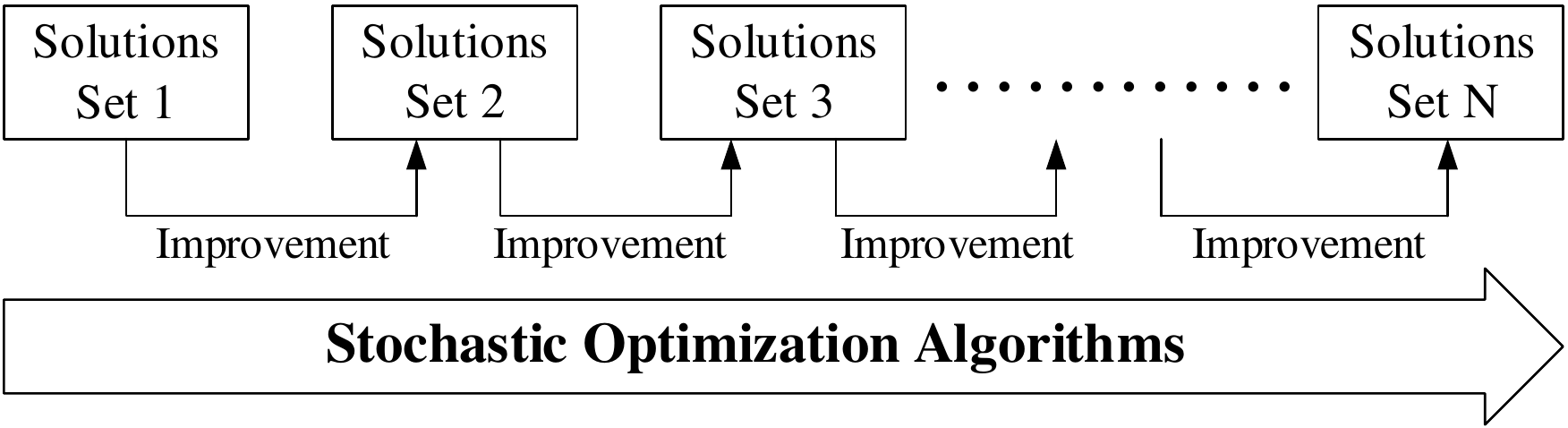}
	\caption{The general framework of stochastic optimization algorithms.}
	\label{fig:stochasitctic}
\end{figure*}

Despite the differences between SI methods and other types of meta-heuristics including evolutionary, physics-based, and event-based ones, they can be considered in the class of stochastic optimization algorithm \cite{faris2018grey}. In this class, sort of randomness involved as opposed to deterministic algorithm, in which there is no random component. Stochastic algorithms start the optimization process with a set of randomly generated solutions. Depending on the underlying mechanism of the algorithm, this set is improved until the satisfaction of an end condition. The set of solutions may have one or more solutions, as shown in Fig.~\ref{fig:stochasitctic}.
The set of solutions is often called the set of candidate solutions since it includes potential solutions for a given problem. If the set has one solution, the algorithm is individual-based or single-solution. On the other hand, the algorithm is called population-based of multi-solution in case of using multiple solutions \cite{faris2018grey}. 

Two questions that naturally arise are about the pros and cons of stochastic and population-based algorithms as compared to deterministic and individual-based ones. Deterministic algorithms find the same solutions in each run and are computationally less expensive. However, they tend to be trapped in locally optimal solutions and mostly require gradient information of the problem. Stochastic algorithms, however, benefit from higher local optima avoidance thanks to the stochastic component. 
On the other hand, single-solution algorithms use one solution in each step of the optimization, and they are computationally cheaper than population-based techniques. Their convergence behavior is also rapid. However, they suffer from entrapment in locally optimal solutions. Population-based algorithms are computationally expensive but better avoid locally optimal solutions. Many studies demonstrate that population-based algorithms can solve complex engineering problems with competitive performance \cite{boussaid2013survey}.  

Due to the advances in the area of population-based stochastic algorithms and the many optimization algorithms, one might ask whether such methods need enhancement. As stated by the No-Free-Lunch theorem \cite{wolpert1997no}, no algorithm is able to solve all optimization problems, that is, one algorithm may perform best in a set of problems but worst in a different set. This theorem has motivated researchers to improve such algorithms significantly. 
One popular method of improving the performance of stochastic algorithms is to hybridize them \cite{talbi2002taxonomy}. A hybrid algorithm can be designed in different ways, including but not limited to, borrowing some of the operators form an algorithm, sequential use of two or more algorithms, adding chaotic maps, etc. The majority of these methods increase the computation cost but result in more accurate results. A taxonomy of hybrid meta-heuristics can be found in \cite{talbi2002taxonomy}.
	
Due to the simplicity of SI methods, gradient-free mechanism, high local optima avoidance, and accuracy in estimating global solutions, they have been widely integrated into systems to solve challenging real-world problems in both science and industry. These characteristics and applicability are often the main reasons why people call such algorithms a black-box optimizer. We do not need to know the internal mechanisms of a problem and estimate optimal solutions by tuning the inputs while monitoring their changes in the output.

The majority of real-world problems also have design constraints that should be considered when evaluating candidate solutions obtained by SI methods \cite{gandomi2012evolutionary}. In case of any violation at any level, the solution is not practical. Such solutions are called infeasible, which means they are optimal but do not satisfy the constraints. There are different methods to handle constraints. Penalty functions are very popular in this area \cite{yeniay2005penalty}, in which a violated solution is penalized so that it does not show better objective values as compared to others. This way, SI methods do not favor them as feasible solutions. 
	
	
There are many swarm algorithms proposed recently. In the following sub-sections, the most well-regarded and recent algorithms in this area are briefly presented. 
	
\subsection{\textcolor{black}{Particle Swarm Optimization}}	
PSO is initially introduced by J. Kennedy and R. Eberhart in 1995   as a global optimization algorithm for nonlinear functions \cite{Kennedy1995Particle}. It mimics the social and individual intelligence of birds in a flock when foraging. This algorithm considers an optimization problem as an $n$-dimensional space, in which $n$ is the number of parameters. Each bird represents a candidate solutions. The birds update their position using the following equation, and the underlying mechanism of search will improve the solutions over time: 	
\begin{align}
& v_{i}^{j}(t+1)=\omega v_{i}^{j}(t)+c_{1}r_{1}(P_{i}^{j}-x_{i}^{j}(t))+c_{2}r_{2}(G_t-x_{i}^{j}(t)), \label{eq:q1} \\
& x_{i}^{j}(t+1)= x_{i}^{j}(t)+v_{i}^{j}(t+1), \label{eq:q2}
\end{align}
	
\noindent where $t$ indicates the iteration number, $\omega$ is a time-varying variable called the inertial weight, $v_{i}^{j}(t)$ shows the current velocity of $j$-th variable in $i$-th particle, $x_{i}^{j}(t)$ is $j$-th variable in $i$-th particle, $r_{1}$ and $r_{2}$ are random numbers in the interval of $ [0,1] $, $c_{1}$ and $c_{2}$ are the individual (cognitive) and social factors, $P$ indicates the best solution obtained by each particle,  and $G_t$ shows the best solution found by all particle from the first to $t$-th iteration. 
	
The first equation shows that particles use their personal best solutions and the best solution obtained by the entire swarm to update their positions. This allows PSO to constantly search around the most promising solutions obtained by the particles while maintaining the best one, which is the most accurate approximation of the global optimum for the problem. 
	
\subsection{\textcolor{black}{Artificial Bee Colony}}
As its name implies, ABC emulates the natural swarming behavior of bees for finding food \cite{karaboga2007powerful}. Just like any other population-based algorithm, there is a population of solutions. This population is divided into three classes: worker beers, onlookers, and scouts. Each worker is associated with a food source and comes back to the nest to the others by means of a waggle dance. Scout bees search for food sources before assigning them to workers. Onlookers choose the food source based on the waggle dance of other bees.
	
In the ABC algorithm, a random initial population of food sources crease and used to calculated new candidate solutions using the following equation \cite{karaboga2007powerful}: 
\begin{equation}
v_{i,j} = x_{i,j} = \phi_{i,j} (x_{i,j}-x_{k,j}),
\end{equation}
where $x_k$ is selected at random from the population with a random variable and $\phi_{i,j}$ takes a random number in [-1,1]. 
	
The first step is considered to create worker bees. They have to now share information about their food sources with the other two types of bees using waggle dance. Onlookers choose food sources using the following equation: 
\begin{equation}
p_i = \frac{fit_i}{\sum_{j=1}^{n} fit_j},
\end{equation}
where $fit$ is the fitness value. 
	
If there is no improvement in any of the food sources, a new solution can be randomized as follows: 
\begin{equation}
x_{i,j} = LB_j + r (UB_j - LB_j),
\end{equation}
where $LB_j$ is the lower bound of the $j$-th variable, $UB_j$ is the upper bound of the $j$-th variable, and $r$ indicates a randomly generated number in the interval of $ [0,1] $.
	
\subsection{\textcolor{black}{Ant Colony Optimization}}
ACO is firstly introduced in 1996 by M. Dorigo \textit{et al.} as a nature-inspired meta-heuristic method for finding the solution for combinatorial optimization problems \cite{dorigo1999ant}. In this algorithm, \emph{stigmergy}, in which communication is done by the manipulation of the environment in swarm, has been used as an inspiration. It has been proved that ants communicate with depositing pheromone on the ground or objects as a way to convey certain messages to other ants. They use different pheromones for different tasks in a colony. For instance, they mark different paths to a source of food from their nest to attract other ants for transportation purposes. Since the longer path faces a longer period of time for vaporization before re-depositing pheromone, the shortest path is chosen automatically using this simple communication. 
	
In the original version of ACO, the problem should be formulated as a graph, each ant is a tour and the objective is to find the shortest one. There are two matrices: distance and pheromone. Each ant uses both of them to update their paths. After several iterations and updating of both matrices, a tour will be established, in which all ants travel through it. This will be considered as the best estimation for the problem. There are other variables of ACO that can be used for problems with continuous variables, constrained, multiple object ices, etc. 
	
\subsection{\textcolor{black}{Firefly Algorithm}}
FA is a meta-heuristic that mimics the mating behaviour of fire flies in nature \cite{yang2010firefly}. This algorithm first creates a population of random solutions. Each firefly in the population can attract others proportional to its fitness value. The main position updating equation is as follows \cite{yang2010firefly}: 
\begin{equation}
x_i = x_i + \beta_0 e^{\gamma r_{ij}}(x_j - x_i) + \alpha \epsilon_i,
\end{equation}
where $\epsilon_i$ is a random number in $ [0,1] $, $x_i$ shows the position of $i$-th solution, and $\gamma$ indicates the light absorption coefficient.

\subsection{Grey Wolf Optimizer}
GWO is invented by Mirjalili \textit{et al.} in 2014 \cite{mirjalili2014grey}, which mimics the social hierarchy of wolves and their hunting behaviors. 
A set of randomly-generated solutions is first initiated and evolved over multiple iterations. During the optimization process, three best solutions obtained so far are used to update the position of others using the following equations \cite{mirjalili2014grey}: 
\begin{equation}
\vec{X}(t+1)=\left(\vec{X}_{1} + \vec{X}_{2} + \vec{X}_3\right)/{3}, \label{eq:gwo4}
\end{equation}
where $\vec{X}_{1}$ and $\vec{X}_{2}$ and $\vec{X}_{3}$ are given as follows:
\begin{equation}
	\begin{split}
	\vec{X}_{1}=\vec{X}_{\alpha}(t)-\vec{A}_{1}\vec{D}_{\alpha},\\
	\vec{X}_{2}=\vec{X}_{\beta}(t)-\vec{A}_{2}\vec{D}_{\beta},\\
	\vec{X}_{3}=\vec{X}_{\delta}(t)-\vec{A}_{3}\vec{D}_{\delta}.
	\end{split}
	\label{eq:gwo1}
\end{equation}
Here, $\vec{D}_{\alpha}$, $\vec{D}_{\beta}$ and $\vec{D}_{\delta}$ are computed as follows:
\begin{equation}
	\begin{split}
	\vec{D}_{\alpha}=\left|\vec{C_{1}} \vec{X}_{\alpha}-\vec{X}\right|,\\
	\vec{D}_{\beta}=\left|\vec{C_{2}} \vec{X_{\beta}}-\vec{X}\right|,\\
	\vec{D}_{\delta}=\left|\vec{C_{3}} \vec{X_{\delta}}-\vec{X}\right|,
	\end{split}
	\label{eq:gwo2}
\end{equation}
where $\alpha$, $\beta$, $\delta$ indicate the notations of three best solutions and stand for three most powerful levels in the hierarchy. 
$\vec{A}$ and $\vec{C}$ are calculated as follows: 
\begin{align}
\vec{A}=2\vec{a} \vec{r}_{1}-\vec{a}, \; \vec{C}=2 \vec{r}_{2}, \label{eq:gwo20}
\end{align}
where $\vec{a}$ linearly decreases from 2 to 0 during the course of run, and  $\vec{r}_{1}$ and $\vec{r}_{2}$ include random numbers in $ [0,1] $. The parameter $\vec{a}$ is a time-varying component, which can be calculated using the following equation: 
\begin{equation}
a = 2 \left(1 - \frac{t}{T} \right),
\end{equation}
where $t$ and $T$ are the iteration index and the maximum number of iterations, respectively.
	
This algorithm has been one of the most widely-used recent meta-heuristics with a wide range of applications that demonstrate the capability of this algorithm in solving real-world problems \cite{faris2018grey}.

\subsection{\textcolor{black}{Whale Optimization Algorithm}}
WOA is proposed in 2016 by Mirjalili and Lewis \cite{mirjalili2016whale}. The WOA is inspired by the bubble net foraging mechanism of humpback whales. This is an intelligent way of catching a school of fish by created and shrinking net of bubbles to drive them towards the surface. 
	
Just like GWO, a set of whales are considered as the candidate solutions for a given optimization problem. Their positions are updated using two mechanisms: encircling prey and spiral movement. The former method is similar to that in GWO using \eqref{eq:gwo1} to \eqref{eq:gwo20}. In the latter approach, a spiral equation is used to update the position. There is a 50\% chance of choosing one of these methods for each of the whales. The mathematical models are as follows \cite{mirjalili2016whale}: 	
\begin{equation} \label{eq:6}
	x(t+1)= 
	\begin{cases}
	x_{p}(t) - AD, & p < 0.5 \\
	D e^{bl} cos(2\pi l) + x_{p}(t),     & p\geq 0.5 
	\end{cases}
\end{equation}
where  $ D=\left | x_{p}(t)-x(t) \right |$ indicates the distance between  $x(t)$ and $x_{p}(t)$, $l$ is a random number in [-1,1], $b$ is a constant that can be tuned, $t$ is the iteration counter, $p$ is a random number in $ [0,1] $ that plays the role of a probability parameter. 
	
\subsection{\textcolor{black}{Harris Hawks Optimizer}}
HHO is invented by Heidari \textit{et al.} in \cite{heidari2019harris}, which mimics the naturally occurring hunting mechanism of Harris hawks. This algorithm uses a number of equations to update the positions of hawks in a search space to mimic different hunting maneuvers that these birds perform to catch preys. The following equation is to provide exploratory behaviour for HHO \cite{heidari2019harris}: 
\begin{equation*} \label{eq:1}
\vec{X}(t+1) = 
\begin{cases}
	\vec{X}_{rand}(t)-r_{1}\left | \vec{X}_{rand}(t)-2r_{2}\vec{X}(t) \right |  & \\  
	\phantom{ccccccccccccccccccccccccccccccccc} \text{for } q \geq 0.5, &  \\ 
	(\vec{X}_{rab}(t)-\vec{X}_{m}(t))-r_{3}(LB+r_{4}(UB-LB)) &  \\
	\phantom{ccccccccccccccccccccccccccccccccc} \text{for } q<0.5, & 
\end{cases}
\end{equation*}
where $\vec{X}(t)$ is the position vector in $t$-the iteration,  $\vec{X}_{rab}(t)$ is the position of the prey, $r_{1}$, $r_{2}$, $r_{3}$, $r_{4}$, and $q$ are random numbers in the interval of $ [0,1] $, $LB$ and $UB$ indicate the lower and upper bounds of parameters, respectively, $\vec{X}_{rand}(t)$ represents a hawk selected randomly from the current population, and $\vec{X}_{m}$ is the average position of the current population of hawks calculated using $\vec{X}_{m}=\frac{1}{N}\sum_{i=1}^{N}\vec{X}_{i}(t)$.

For exploitation, this algorithm uses soft besiege and hard besiege as follows: 	
\begin{align} 
& \vec{X}(t+1)=\Delta \vec{X}(t)-E\left |J\vec{X}_{rab}(t)-\vec{X}(t)\right |, \label{eq:4} \\
& \Delta {X}(t)=\vec{X}_{rab}(t)-\vec{X}(t), \label{eq:deltaX} \\ 
& \vec{X}(t+1)=\vec{X}_{rab}(t)-E \left |\Delta \vec{X}(t) \right |, \label{eq:hard} 
\end{align}
where $\Delta {X}(t)$ is the distance between the solution and the prey in $t$-th iteration, $r_{5}$ is another random variable in the interval of $ [0,1] $, and $J$ changes randomly. 

\subsection{Summary}
In this section, we have reviewed the fundamentals of SI and several well-known SI techniques. Thanks to its superiority and particularities, SI has been used for the design and optimization in many engineering domains.
In the following sections, we review the state of the art applications of SI for NGN. In particular, we focus on three main themes: spectrum management and resource allocation, wireless caching and edge computing, and network security, and also a number of miscellaneous yet important issues.

\section{Spectrum Management and Resource Allocation}
\label{Sec:SpectrumManagement}

\begin{figure*}[t]
	\centering
	\includegraphics[width=0.75\linewidth]{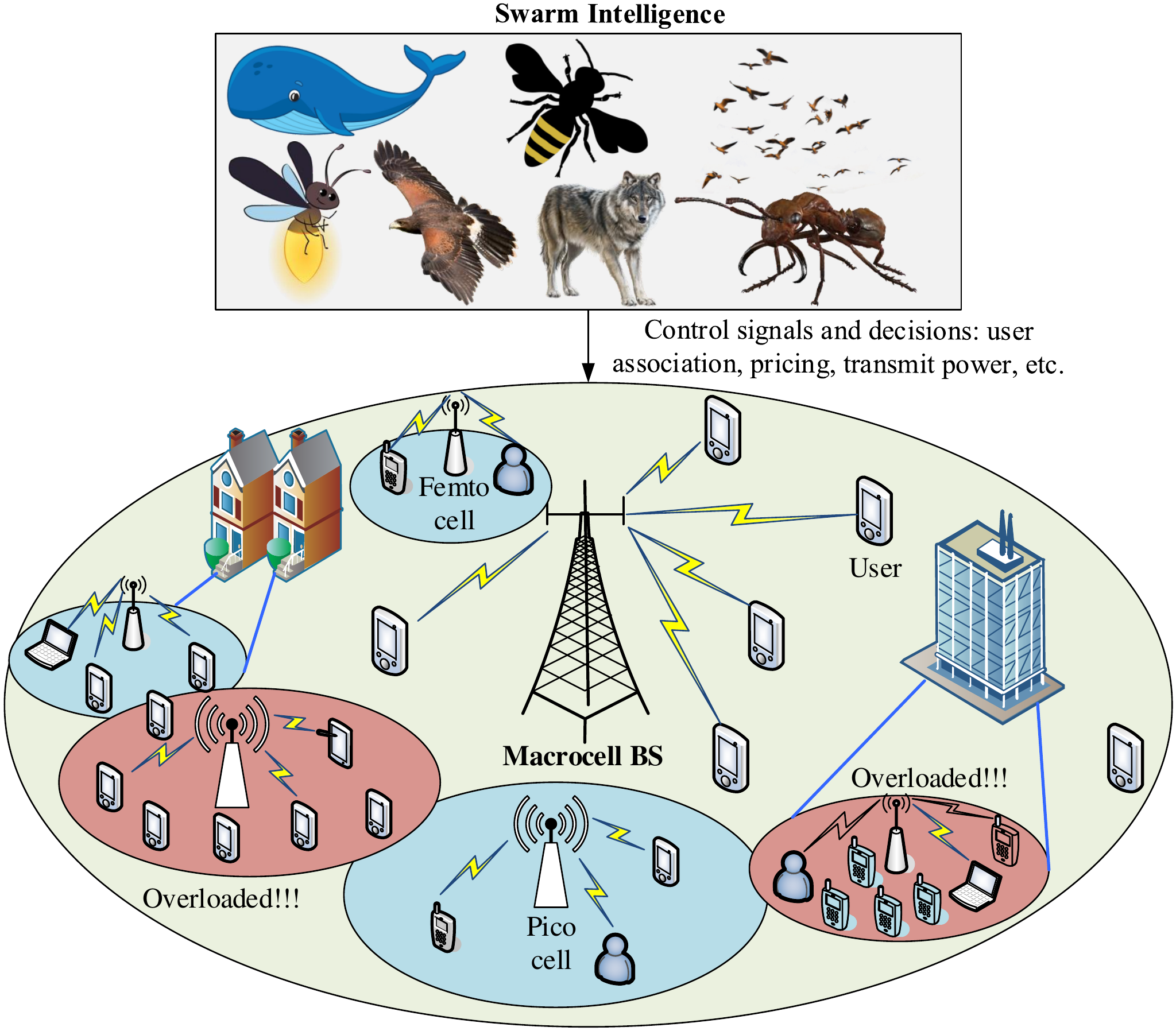}
	\caption{Illustration of a three-tier HetNet, including macro cell, pico cell, and femto cell. Users may compete with the others to associate with their desired BSs and each BS should control the load for better load balancing.}
	\label{Fig:SI_HetNets}
\end{figure*}

Spectrum management (also known as spectrum sharing) and resource allocation are two key techniques to achieve the key 5G performances (1000-fold for throughput, 10-fold for spectrum efficiency, and 100-fold for energy efficiency). Three aspects in spectrum sharing are identified \cite{Zhang2018SpectrumSharing}: 1) renewing the network architecture, e.g., more small cells, 2) exploring more spectrum resource, e.g., millimeter-wave (mmWave) and unlicensed spectrum, and 3) adopting new communication techniques, e.g., cognitive radio networks, D2D, NOMA, Long-Term Evolution (LTE) in unlicensed spectrum (LTE-Unlicensed, LTE-U), and full-duplex communication. 
While the purposes of the first and second ones are to achieve higher spatial reuse gains and more spectrum bands, the third one is to achieve higher spectrum efficiency. 
Besides spectrum sharing, resource allocation is of paramount importance to improve the performance of 5G wireless networks. In general, 5G is expected to support communications, computing, content delivery, and control, and lots of resource types need to be considered \cite{Pham2020Survey_MEC, Mao2017_aSurveyMEC}. Simple examples of resource allocation are user clustering and power control in NOMA/D2D/full-duplex systems, user association and power allocation/load balancing in heterogeneous networks (HetNets), and beamforming optimization in MIMO systems. Resource allocation represents a central theme of this section. 

\subsection{\textcolor{black}{Power Control and Channel Allocation}}
The power and channel allocation problems play a major role in various wireless systems. The term ``channel allocation" varies in different network scenarios, e.g., user clustering/grouping in NOMA, channel/subcarrier allocation in D2D communications and HetNets. For instance, the spectrum allocated to the cellular users can be re-used by one or more D2D pairs (a D2D transmitter and a D2D receiver) on the condition that the interference level imposed by D2D users to the co-channel cellular users is smaller than a predefined threshold. In such scenarios, natural questions are how to adjust the transmit power of D2D transmitters and how to allocate the channels to D2D users. Another example can be the optimization of user clustering and power allocation (among different clusters and within a cluster) in NOMA systems, and subcarrier allocation and power control in HetNets. Various performance metrics would be obtained, e.g., spectral and energy efficiency, secrecy rate, minimal outage probability, etc. The main focus of this subsection is on the optimization of power control and channel allocation in 5G wireless systems using SI algorithms. Fig.~\ref{Fig:SI_HetNets} illustrates the diagram of a three-tier HetNet (macro cell, pico cell, and femto cell) and applications of SI. 

A number of problems in D2D communications (e.g., interference management, resource allocation, and power control) have been solved by SI techniques. Two joint channel allocation and power control problems are studied in \cite{xu2018joint}, where the channel occupied by one cellular user can be reused by multiple D2D pairs. While the first problem considers individual rate requirements for cellular users, the second problem requires the total rate of all cellular users to be greater than a threshold value. For both problems, the standard continuous and binary PSO schemes are used to update power allocation and channel allocation, respectively. The proposed algorithm is then compared with a random PSO scheme, where channels for D2D pairs are randomly assigned. This work is extended to the unlicensed band in \cite{girmay2019joint}, where D2D pairs causing strong interference to cellular users are forced to use the unlicensed band. Simulations demonstrate that the proposed approach outperforms the algorithm in \cite{xu2018joint} (only licensed band) and the scheme when only one D2D pair is allowed to reuse the same licensed channel. It is elaborated in \cite{Sun2019Coalition} that coalition game cannot achieve high performance and low complexity simultaneously. To solve this challenge, the authors consider the concept of ``\textit{priority sequences}" to improve the initialization and formation phases of the coalition game. To further improve the D2D sum rate, the WOA is used to solve the power control problem. In particular, the objective function is designed in a such way that maximizes the D2D sum rate while guaranteeing minimum rates of cellular users.

Considering as a potential multiple access technology for NGN, NOMA has received much interest from the research community. Besides game theory (e.g., matching game and coalitional game) and machine learning, SI is a vital alternative for solving user clustering and power control problems in NOMA systems. The work in \cite{masaracchia2019pso} adopts the PSO to minimize the total transmit power by optimizing NOMA pairing decisions. Then, two representative scenarios are considered: disaster and UAV communications, where low power consumption is an important aspect. An interesting observation from this work is that minimizing the transmit power depends on not only channel conditions, but also the number of sub-channels as well as the available bandwidth. Another application of PSO can be found in \cite{jiao2020network} for NOMA resource allocation in satellite-based IoT systems. The Lyapunov optimization framework is adopted to transform the long-term problem of data and power allocation into a sequence of online problems, which are then solved at the corresponding time slots. Compared with the schemes using the Karush–Kuhn–Tucker (KKT) optimality conditions, the PSO-based scheme achieves better utility with a lower average delay, while obtaining a good tradeoff between complexity and storage. Very recently, the HHO is utilized to solve the joint UAV placement and power allocation problem for NOMA-enabled visible light communications \cite{pham2020sum}. The main advantage of HHO over existing methods is that the problem can be solved directly instead of solving two subproblems iteratively in an alternative manner. Moreover, the HHO-based algorithm relaxes the assumption of exponential-distance  models in existing studies. Simulation results show that the HHO-based algorithm is superior to the conventional orthogonal multiple access (OMA), power allocation, and random schemes under various settings, and also outperforms several SI techniques such as PSO, evolution strategy, and genetic algorithm (GA). A combination of D2D and NOMA in MEC systems is studied in \cite{diao2019joint}. In particular, the joint computing resource, power and channel allocation problem is investigated and the PSO is applied to allocate transmit power within individual groups and for different groups. Other applications of SI techniques for edge computing can be found in various studies, e.g., joint resource allocation in dense HetNets \cite{guo2018efficient} and in multi-tier MEC systems \cite{huynh2020efficient}.

\subsection{\textcolor{black}{User Association And Load Balancing}}
User association refers to the process of associating users with their preferred BSs (also known as eNB and gNB in 5G and sixth-generation (6G), respectively), and load balancing refers to the distribution of workloads across multiple BSs. Over the last few years, many studies on user association and load balancing in HetNets, multi-antenna systems, and wireless-powered communication networks have been conducted. The excellent surveys on user association and load balancing can be found in \cite{Liu2016UserAssociation} and \cite{Andrews2014anOverview_LoadBalancing}, respectively. According to these surveys, five metrics are mainly used for user association, including outage and coverage probability, spectrum efficiency, energy efficiency, quality of service (QoS), and fairness. Moreover, the frameworks from convex optimization, game theory, and stochastic geometry are used to provide tractable methods for optimization problems of these issues. Henceforth, we focus on reviewing user association and load balancing mechanisms, which are designed with the help of SI techniques. 

The PSO is adopted in \cite{kuribayashi2020particle} to overcome drawbacks of the conventional cell range expansion techniques, i.e., users tend to connect to the BS with the highest received signal power. Instead of maximizing the sum rate as considered in existing literature, this work considers maximizing the number of users with satisfactory rate requirements and the number of BSs with at least one connected user. The PSO is used to optimize the bias values assigned to individual BSs, thus implicitly improving the load balancing in dense HetNets. Experiments illustrate that the PSO-based algorithm can achieve superior performance (in terms of resource utilization and the ratio of satisfactory users) than the unified cell range expansion bias approach and classical PSO methods. 
Since the limited resources of small cell BSs and the QoS of users are not jointly considered in existing user association mechanisms, the work in \cite{zhong2019stable} proposes to use a multi-leader multi-follower Stackelberg game to solve this issue in backhaul-limited HetNets. In this context, BSs are regarded as the first mover and can adjust the prices according to their loads, while users are considered as the follower and compete with each other to have the desired association. The PSO is adopted at the BSs to determine their prices and obtain the perfect Nash equilibrium. 
The work in \cite{plachy2019joint} considers a joint BS's position and user association in a dense network, where UAVs can be deployed as flying BSs. Two evolutionary algorithms, PSO and GA, are then applied to obtain the optimal UAV's positions. Both the PSO- and GA-based algorithms are shown to outperform the k-means clustering scheme. Interestingly, there is a trade-off between PSO and GA, while the PSO-based algorithm has lower complexity, the GA-based scheme achieves better network throughput, radio of satisfactory users, and active number of flying BSs. Despite a promising candidate for high rate improvements, terahertz (THz) ultra-dense HetNets require novel user association mechanisms that consider unique features (e.g., highly directional transmission and noise-limited environment) of THz channels \cite{Boulogeorgos2018Users}. This work takes into account these particularities of ultra-dense THz networks and proposes to use the GWO to devise a novel user association strategy. Experiments show that the GWO-based algorithm is superior to the PSO-based approach in terms of convergence and system rate. Applications of the WOA techniques for user association and task offloading decision in edge computing systems are considered in \cite{pham2020whale}. In \cite{Jiang2020Deep}, the PSO is utilized to solve the user association problem and generates a training dataset for a deep neural network. To avoid being struck at local solutions and accelerate the convergence rate, this work also proposes a novel probability metric for user association that considers small-scale fading and interference values. In the context of multi radio access technologies, game theory and the ACO are adopted to address the user association problem in \cite{Li2018Context}.

Besides user association in ultra-dense HetNets, SI techniques have found lots of applications for load balancing in cloud and edge computing. To overcome the issue of unbalanced workload in mixed manufacturing, an energy-aware load balancing and job scheduling is studied in \cite{wan2018fog}. Fog nodes are used to abstract and then an energy consumption model of smart factory equipment, whereas the PSO is employed to find the workload allocation solution. An adaptive-step size algorithm based on the glowworm swarm optimization (GSO) and sine cosine algorithm (SCA) is proposed in \cite{dong2019joint} to find the server selection and improve the load balancing between edge servers and the central cloud. For mobile learning services, a dominant FA approach is proposed for load balancing among cloud virtual machines in \cite{sekaran2019improving}. The proposed dominant FA algorithm can ensure that learners can receive m-learning content without delay, and interestingly the proposed algorithm is superior to various methods, including ACB, ACO, and the weighted round robin algorithm. The GWO and PSO are jointly used to find the solution of load balancing in cloud computing \cite{Gohil2018AHybrid}. Another work on load balancing in cloud computing is \cite{ebadifard2018pso}, where the PSO is used to distribute the tasks among cloud virtual machines. 

\subsection{\textcolor{black}{MIMO Detection, Channel Estimation, and Precoding}}
Regarding the network model presented in this subsection, we consider an MIMO system with $ N_{t} $ and $ N_{r} $ antennas at the transmitter and the receiver, respectively. The received signal vector is given as 
\begin{equation}\label{Eq:complex_rs}
\mathbf{y} = \mathbf{H} \mathbf{x} + \mathbf{n},
\end{equation}
where $ \mathbf{H} \in \mathbb{C}^{N_{r} \times N_{t}} $ is a matrix that represents the channel between two antenna (each one at the receiver and transmitter), $ \mathbf{x} \in \mathbb{C}^{N_{t}} $ is a vector of symbols drawn from a complex constellation, $ \mathbf{n} \in \mathbb{C}^{N_{r}} $ is vector of complex noise. 

\subsubsection{\textcolor{black}{MIMO Detection}}
For many years, MIMO has been a key enabler of high spectral efficiency, coverage range, and reliability without the need for additional spectrum. In MIMO, the receiver may use some detection strategies to estimate the transmitted symbols that may be corrupted by interference and noise over the medium. However, the complexity of a detection mechanism highly depends on the system size (e.g., the numbers of transmit and receive antennas, and the modulation scheme), matrix multiplication and inversion operations \cite{Albreem2019MassiveMIMO}. 

An MIMO detection mechanism is to estimate the transmitted vector $ \mathbf{x} $ based on the received signals and the channel matrix. The maximum likelihood is an optimal algorithm when the probability of error is used as the optimization metric. A prerequisite step to apply SI techniques for MIMO detection is transforming the complex-valued representation~(\ref{Eq:complex_rs}) into a real-valued expression. As a result, the signal dimension is double, i.e., one for the real part and another for the imaginary part. Therefore, it is assumed that SI techniques are adopted for the real-valued expression. The MIMO signal detector tries to exhaustively search over all the possible signals (i.e., $ \mathbf{x} \in \mathcal{X}^{2 \times N_{t}} $) to solve the following problem:
\begin{equation}\label{Eq:CostFunction_Symbol}
\hat{\mathbf{x}}(\mathbf{y}) = \argmin_{\mathbf{x} \in \mathcal{X}^{2 \times N_{t}}} || \mathbf{y} - \mathbf{H} \mathbf{x} ||^{2},
\end{equation}
where $ \mathcal{X} $ is the set of real-valued entries in the signal constellation, for example, $ \mathcal{X}=\{-3,-1,1,3\} $ for 16-quadrature amplitude modulation (QAM) and $ \mathcal{X}=\{-7,-5,-3,-1,1,3,5,7\} $ for 64-QAM. 
Unfortunately, the problem involves a search over a very huge solution space and the complexity of the maximum likelihood method would increase exponentially with respect to the number of transmitter antennas. For instance, to obtain the optimal signal, the detector needs to perform $ 16^{8} \approx 4.295 \times 10^{9}$ enumerations for an MIMO system with $ 8 $ antennas at the transmitter to support 16-QAM. Hence, it is almost impossible to use the maximum likelihood detector in MIMO systems. In such a case, the use of MIMO linear/non-linear detectors (e.g., zero forcing (ZF), best linear estimator (BLE), successive interference cancellation (SIC) or vertical Bell laboratories layered space-time (V-BLAST), and sphere decoding (SD) are beneficial. However, they have a critical drawback due to the matrix inversion and QR decomposition that incur a high computational complexity. Recently, many detection mechanisms based on approximate inversion methods, local search, belief propagation, and machine learning have been proposed \cite{Albreem2019MassiveMIMO}. In this part, we focus on detection algorithms using SI techniques for MIMO systems. 

Thanks to their advantages, SI algorithms can be considered as alternative approaches for MIMO detection problems. The first application of the ACO for the MIMO detection problem is proposed in \cite{Lain2010NearMLD}. Inspired by the ACO algorithm, the MIMO maximum likelihood detection (MLD) can be formulated as the shortest path problem, in which each transmit antenna is regarded as a city and the transmitted symbols are viewed as the potential paths to a specific city. Corresponding to a solution component in the ACO, a similar concept of path movement is defined in \cite{Lain2010NearMLD}. The solution component $ s_{ij} $ indicates the solution that the antenna $ i $ should transmit the symbol $ j $, where $ j \in \{1, \dots, |\mathcal{X}|\} $. Indeed, the cardinality value $ |\mathcal{X}| $ depends on the QAM type, e.g., $ |\mathcal{X}| = 2 $ for 4-QAM, $ |\mathcal{X}| = 4 $ for 16-QAM, and $ |\mathcal{X}| = 8 $ for 64-QAM. The cost function to estimate transmitted symbols is evaluated in~\eqref{Eq:CostFunction_Symbol}, which is then converted to a heuristic value by means of a sigmoid function. Next, the pheromone levels and Euclidean distance values are used to calculate the transition probabilities, which are used to select the transmitted symbols. To be more specific, the symbol with a smaller distance has a higher chance to be selected by the artificial ant. To avoid the convergence to a local optima in the original ACO, the authors in \cite{Lain2010NearMLD} adopt a modified ACO scheme, in which ants are enabled to have non-identical sets of pheromone trails, i.e., the hard decision symbol used in estimating the distance is now replaced by the tentatively chosen symbol. From the simulations, the ACO-based approach in \cite{Lain2010NearMLD} can optimize the tradeoff between the computational complexity and symbol error rate (SER) performance. In particular, for a 16-QAM 16$ \times $16 MIMO system, the ACO can achieve a near-MLD SER performance when the number of ants is around $ 20,000 $. 

The work in \cite{Lain2010NearMLD} can be considered as a pioneering study on SI for the MIMO detection problem. However, the advantage of the proposed ACO algorithm in \cite{Lain2010NearMLD} is limited by a large number of ants and slow convergence. A hybrid algorithm, as a combination of the ACO and PSO, is proposed in \cite{Mandloi2016aLowComplexity}. Unlike the ACO and PSO based MIMO detectors, the concepts of distance in ACO and velocity in PSO are jointly considered to create a new probability metric, i.e., the fitness value is calculated as a weighted sum of the distances traveled by the solutions (i.e., particles) and the velocity of each solution. The proposed hybrid algorithm is stated to offer several advantages compared to the original ACO and PSO: 1) avoiding convergence to locally optimal solutions, 2) faster convergence, 3) lower computational complexity, and 4) removing the sensitivity to the choice of initial solutions as in \cite{Adnan2007PSO_SybmolsDetection}. The simulation results in \cite{Mandloi2016aLowComplexity} show that the proposed scheme can perform close to the SD detector and is superior to the existing ACO detection mechanisms in \cite{Lain2010NearMLD, Mandloi2015CongestionControl}. More specifically, at the target BER of $ 10^{-3} $, the proposed hybrid algorithm attains $ 2.5 $ dB and $ 4 $ dB improvement in signal-to-noise ratio with respect to the ACO algorithms in \cite{Lain2010NearMLD} and \cite{Mandloi2015CongestionControl}, respectively. More recently, the FA algorithm is used to investigate a robust detection in \cite{Datta2019aNearML}; however, the detector developed in this work is applicable for MIMO systems with 4-QAM signaling only. For future work, detection schemes for massive MIMO systems with higher modulation orders can be investigated. 

\subsubsection{\textcolor{black}{MIMO Beamforming}}
Beamforming plays a major role in MIMO systems because the transmitter needs to direct the spatial data-stream at the intended receiver. Basically, beamforming implies that multiple antennas are used to adjust the wave directions by appropriately changing the magnitude and phase of each antenna in the array. Terms beamforming and precoding are usually used interchangeably in the literature. 
There are three types of beamforming in multi-antenna wireless systems: digital beamforming, analog beamforming, and hybrid digital-analog beamforming. The use of fully digital beamforming is prohibitively complex in massive MIMO systems because of the number of radio frequency (RF) chains, whereas the performance of MIMO systems with fully analog beamforming would reduce due to the effect of phase shift quantization and the lack of amplitude adjustment \cite{Zeng2019EnergyEfficient}. Compared with the first two kinds, hybrid beamforming can achieve good performance by appropriately adjusting the number of RF chains \cite{Molisch2017HybridBeamforming}. The beamforming matrices are typically optimized using optimization and game-theoretic approaches. Thanks to their advantages, SI and bio-inspired techniques are promising candidates to solve beamforming optimization problems, which represents the theme of this subsection. 

The work in \cite{Hefnawi2016LargeScale} considers the concept of cooperative MIMO in cognitive radio enabled WSNs. In particular, a very large number of sensors are divided into different clusters and each cluster is treated as a virtual antenna in a multi-cluster MIMO system. The PSO approach is applied to optimize the transmit beamforming vectors so as to maximize the ergodic capacity of each cluster. The main benefit is that the PSO-based method does not require gradient search operations, which can reduce the complexity significantly. Simulation results show that the PSO-based scheme can achieve competitive system capacity, and interestingly impressive performance can be obtained when the number of clusters is small. Owing to the importance of spectrum sensing in cognitive radio networks, various SI-inspired methods to improve spectrum sensing have been proposed in the last few years. For example, a hybrid PSO-gravitational search algorithm (GSA) is studied in \cite{Eappen2020HybridPSO_GSA} for optimizing the performance of joint spectrum sensing and energy efficiency, and a random walk GWO method is investigated in \cite{Karthikeyan2019Guided} for deriving sensing and data transmission schedules.

To enable low-complexity and efficient mmWave massive MIMO communications, various studies have been dedicated to beamforming optimization. The work in \cite{Zhu2019JointTxRx} considers to jointly optimize transmit (Tx) and receive (Rx) beamforming vectors and power allocation for a NOMA-mmWave MIMO system. Since the the sub-problem of Tx-beamforming is non-convex and the original PSO may result in a locally optimal solution, a boundary-compressed PSO approach is devised. In particular, two boundary values, inner and outer, are introduced, and the inner value increases linearly during the course of optimization. It is shown that compared with the conventional OMA and 2-user downlink case, the proposed boundary-compressed PSO can achieve a higher sum rate with reasonable convergence rate and stability. A similar study can be found in \cite{Xiao2019UserFairness}, where the max-min problem is optimized and the PSO is used to obtain the upper bound value. Also using the boundary-compressed PSO, the same authors extend \cite{Zhu2019JointTxRx} to consider hybrid beamforming in \cite{Zhu2019MillimeterWave} and show that the proposed algorithm is superior to the conventional OMA and fully digital beamforming cases in terms of spectral efficiency and energy efficiency, respectively. To improve the robustness of hybrid beamforming schemes, the work in \cite{almagboul2019efficient} considers improving the bat algorithm to mitigate interference and estimate directions of arrival in the digital beamforming domain. Experiments illustrate that the proposed bat algorithm achieves good stability and outperforms the original one and PSO approaches.

SI approaches are also considered to solve various optimization problems in MIMO systems. To obtain a tradeoff between spectral efficiency and energy efficiency for MIMO systems, the work in \cite{nimmagadda2020optimal} proposes two SI approaches, namely improved random vector-based GWO and improved random vector-based lion algorithm (LA). The transmit power and beamforming vectors are optimized to maximize a multi-objective function of spectral and energy efficiencies. The simulation shows that the proposed SI approaches can obtain competitive performance and outperforms several SI methods such as PSO, FA, GWO, and LA. Recently, the work in \cite{souto2020beamforming} considers an MIMO system that is assisted by a reconfigurable intelligent surface (RIS). In general, an RIS is composed of many passive elements that can reflect incident signals by adjusting their phases and amplitudes. However, estimating CSI accurately in RIS-aided MIMO systems is a challenging task, and state-of-the-art solutions may not be practical because of large overheads, high complexity, and hardware challenges \cite{souto2020beamforming}. To solve these challenges of channel estimation, the authors in \cite{souto2020beamforming} adopt the PSO to optimize active beamforming at the BS and passive beamforming at the RIS. Unlike conventional methods, the PSO-based method is not dependent on CSI acquisition and does not require one RF chain for each RIS passive element, thus reducing the cost and energy consumption.

\subsubsection{\textcolor{black}{MIMO Channel Estimation}}
Channel estimation has been considered a vital means to enhance the overall performance of wireless systems. With accurate channel estimation, the potential advantages of massive MIMO systems (and other networks as well) can be exploited to enhance the network performance. Nevertheless, there are several challenges related to channel estimation problems, e.g., time variance and frequency selectivity of wireless channels, pilot contamination among adjacent cells, a large number of channels (e.g., in an intelligent reflecting surface (IRS) based systems), etc. In this part, we focus on channel estimation techniques using SI algorithms. For more details on channel estimation, we suggest interested readers to read the survey paper \cite{Liu2014ChannelEstimation}.

\begin{table*}[t]
	\caption{Summary of the state-of-the-art SI techniques for spectrum management and resource allocation.}
	\label{Tab:Summary_Table_III}
    \resizebox{\textwidth}{!}{	
	\begin{tabular}{|c|c|c|p{12.5cm}|}
		\hline 
		Category & Paper  & Techniques & Highlights  \\ 
		\hline
		\hline
		
		\multirow{8}{*}{\makecell{Power control \\and \\channel allocation}} & \multirow{2}{*}{\cite{xu2018joint, girmay2019joint}}  & \multirow{2}{*}{PSO} & Each cellular user can share its occupied channels with multiple D2D pairs. Extension to consider both licensed and unlicensed (WiFi) bands in \cite{girmay2019joint} would significantly improve the system rate. \\
		
		\cline{2-4} 
		
		& \multirow{3}{*}{\cite{Sun2019Coalition}} & \multirow{3}{*}{WOA} & An improved variant of coalitional game is used for channel allocation. \\
		&&& The WOA is used for power allocation and the fitness function is designed to maximize the total sum rate of D2D users while guaranteeing the QoS requirements of cellular users. \\ 
		
		\cline{2-4} 
		
		& \multirow{4}{*}{\cite{pham2020sum}} & \multirow{4}{*}{HHO} & The HHO is used to solve a joint UAV placement and power allocation problem in NOMA-based visible light communications. The HHO is also employed as a trainer of deep neural networks. \\
		
		&&& The HHO-based algorithm is superior to the conventional OMA and an existing power allocation scheme, and outperforms other metaheuristics: PSO, evolution strategy, and genetic algorithm.\\
		
		\hline	
		
		\multirow{8}{*}{\makecell{User association \\and \\load balancing}} & \multirow{2}{*}{\cite{kuribayashi2020particle}}. & \multirow{2}{*}{PSO} & The fitness function is to maximize the numbers of satisfactory users and serving BSs. The PSO-based schemes achieve better performance than conventional user association mechanisms.  \\
		
		\cline{2-4} 
		
		& \multirow{2}{*}{\cite{Boulogeorgos2018Users}} & \multirow{2}{*}{GWO} & Unique features of THz bands are considered for user association. The GWO converges faster and yields a higher sum rate than the PSO-based approach. \\ 
		
		\cline{2-4} 
		
		& \multirow{1}{*}{\cite{wan2018fog}} & \multirow{1}{*}{PSO}  &  Energy-aware load balancing in fog computing enabled smart manufacturing is optimized by the PSO. \\ 
		
		\cline{2-4} 
		
		& \multirow{2}{*}{\cite{sekaran2019improving}} & \multirow{2}{*}{FA}  & A dominant FA approach is proposed for load balancing among cloud virtual machines. Thanks to the proposed FA algorithm, mobile learning services are provided without delay. \\ 
		
		\hline
		
		\multirow{12}{*}{\makecell{MIMO \\applications}} & \multirow{3}{*}{\cite{Lain2010NearMLD}} & \multirow{3}{*}{ACO} &  The MIMO detection problem is formulated as the shorted path optimization and then solved by the ACO. \\
		&&& A tradeoff between complexity and SER is obtained, and the ACO-based scheme shall approach the near-MLD performance when the number of ants is sufficiently large. \\
		
		\cline{2-4} 
		
		& \multirow{2}{*}{\cite{Mandloi2016aLowComplexity}} & \multirow{2}{*}{\makecell{PSO \\and ACO}} &  A hybrid (PSO and ACO) scheme is adopted for MIMO detection and can achieve a gain of 2.5 dB over the ACO in \cite{Lain2010NearMLD}. \\ 
		
		\cline{2-4} 
		
		& \multirow{3}{*}{\cite{Hefnawi2016LargeScale}} & \multirow{3}{*}{PSO} &  The PSO is used to optimize transmit beamforming of virtual antenna, each is comprised of multiple sensor nodes. Besides system rates, the computational complexity is significantly reduced when compared with the conventional gradient-based methods. \\
		
		\cline{2-4} 
		
		& \multirow{2}{*}{\cite{nimmagadda2020optimal}} & \multirow{2}{*}{\makecell{GWO \\and LA}} &  A tradeoff between spectral and energy efficiencies is optimized and significant improvements are reported over PSO, FA, and original GWO and LA. \\
		
		\cline{2-4} 
		
		& \multirow{2}{*}{\cite{jiang2019joint}} & \multirow{2}{*}{WOA} & Three improved versions of the WOA and compressed sensing are jointly employed for pilot allocation in acoustic OFDM systems. \\
				
		\hline
	\end{tabular}
    }
\end{table*}

Among SI techniques, PSO is a vital alternative for well-known channel estimators such as minimum mean squared error (MMSE) and least-squares (LS). In \cite{knievel2012particle}, PSO and cooperative PSO (CPSO) are applied for channel estimation in terms of mean squared error (MSE) performance. The CPSO is highly preferable when the search space has a very high dimension since the whole search space is partitioned into smaller subsets, and multiple swarms (i.e., solutions) are maintained at a time instance instead only one swarm in the PSO. Thus, the CPSO is suitable for massive MIMO systems, where massive antennas are used at the transmitter. The channel estimation problem becomes more difficult when it couples with other optimization variables. For example, the work in \cite{zhang2013evolutionary} investigates a joint channel estimation and multi-user detection problem, which is solved by four evolutionary techniques (including genetic algorithm, repeated weighted boosting search, PSO, and differential evolution). In particular, an iterative algorithm is designed by applying continuous versions for channel estimation and binary versions for multi-user detection. Recently, the work in \cite{jiang2019joint} applies an enhanced variant of the WOA and compressed sensing to allocate the pilot in underwater acoustic OFDM systems. To improve the original WOA, three strategies are investigated, including: 1) good initialization to avoid the prematurity issue, 2) chaotic switching to prevent traps from falling into a local minimum, and 3) nonlinear parameters to appropriately adjust the transition between exploration and exploitation. The proposed enhanced WOA is shown to have better performance than some other algorithms such as PSO, original WOA, and GA. Similar to \cite{jiang2019joint}, some other SI techniques are also used for channel estimation in sparse networks, for example, GWO in \cite{csimcsir2017pilot} and FA in \cite{tacspinar2018efficient}.

\subsection{Summary}
In this section, we survey the applications of SI for spectrum management and resource allocation in NGN. In particular, we discuss three aspects, including power control and channel allocation, user association and load balancing, and MIMO applications (signal detection, channel estimation, and precoding). There are numerous advantages in using SI techniques for optimization of spectrum management and resource allocation, e.g., obtaining high-quality solutions, generating datasets as the input of deep neural networks, and solving challenging problems that are difficult to solve by other methods like convex optimization and game theory. It can be observed that the PSO is widely used for many problems, while other SI techniques may find their suitability and practicality in several other applications. For example, since the original ACO is proposed to solve mixed-integer problems, it is highly suitable for MIMO signal detection. Another observation is that as each SI approach has its distinct features and disadvantages, combining multiple SI techniques to solve a specific problem is widely investigated and the combined algorithm typically outperforms individual algorithms. Besides, recent SI techniques have demonstrated better performance than the old ones, but the comparison among recent advances is usually not available in the literature. In summarizing this section, the reviewed literature, used techniques, and highlights are summarized in Table~\ref{Tab:Summary_Table_III}.
\section{Wireless Caching and Edge Computing}
\label{Sec:WirelessCaching}
One of the important missions of 5G wireless networks is distinctly different and complex to previous wireless system generations. It is mainly due to the proliferation of massive connectivity and the emergence of many new applications, which are usually compute-intensive, latency-critical, energy-consuming, and bandwidth-hungry. A white paper from Cisco \cite{Cisco2017VNI} shows that there will be around $ 28.5 $ billion connected things in 2022, up from $ 21.5 $ billion devices in 2019, and the amount of traffic will increase very fast, e.g.,  $ 12 $-fold for augmented/virtual reality (AR/VR) and ninefold for Internet gaming. Although many technologies (e.g., MIMO, millimeter-wave (mmWave) communication, HetNets) have been introduced to meet the traffic requirement in NGN, the deployment and expenditure obstacles of wired and wireless backhaul links have motivated another solution. In this case, caching at the wireless edge has emerged as a great solution \cite{liu2016caching}. Moreover, a mobile equipment typically has limited computing capability, storage, and battery capacity to run emerging applications and services (e.g., mobile blockchain, AR/VR, real-time online gaming). To deal with this challenge, the concepts of MEC and fog computing have been introduced by the European Telecommunications Standards Institute (ETSI) Industry Specification Group (ISG) in 2014 and by Cisco in 2012, respectively. Conceptually, MEC and fog computing move the IT functionalities, storage, and computing capabilities from central clouds to the network edge, enabling these computing paradigms implemented on new compute-intensive and latency-critical applications at the device level. In MEC systems, a critical use case from the user perspective is computation offloading and its joint optimization with communication, caching, and control. 

\subsection{Wireless Caching}
\label{SubSec:WirelessCaching}
Back to several decades ago, caching has been developed for various application domains such as web caching and memory caching in operating systems. As of massive traffic, user connection, and network densification, and stringent services in NGN, the centralized caching cannot accommodate user demands \cite{liu2016caching}. The reasons are mainly from the large latency, limited backhaul capacity, and low reliability. Therefore, over the past few years, the concepts of wireless caching have been studied in a number of investigative projects. Generally, content placement and content delivery are the two main problems in wireless caching. While the first problem decides the location and size of each content chunk, the latter problem is about how to deliver the content to the requesters \cite{liu2016caching}. Indeed, SI techniques have found their applications in achieving competitive performance for wireless caching systems.

In wireless caching, caching at the edge is the main theme and has been considered in many studies. For example, the work in \cite{Li2019SociallyAware} proposes a socially aware caching mechanism in D2D-enabled for radio access networks. The proposed scheme is composed of three main phases, in which the first phase deals with the clustering of fog nodes, each one is typically located at a corresponding BS. After the clustering is completed, a center user scheme is proposed to select users within the coverage of each fog node to cache data for the other users. After that, the caching problem at the selected users is formulated as a discrete problem and then solved by the ACO. Numerical results show that the ACO-based approach can outperform several baseline schemes in terms of cache hit ratio and average total delay. Applications of SI techniques (PSO and bat algorithm) are also investigated in \cite{Naguib2018Group, ali2019optimized} for content caching in D2D communications. 
Another work on wireless caching at the edge is considered in \cite{Jin2019IVCN}. In particular, a new concept, namely information-centric virtual content network, is proposed by combining network function virtualization and content centric network. Based on the proposed architecture, an optimization problem is studied to minimize the total weighted hops in a HetNet. Similar to \cite{Li2019SociallyAware}, the problem is also solved by the ACO and then compared with several baseline caching mechanisms like cache everything everywhere and popularity-based schemes. Various caching mechanisms in vehicular networks are studied in \cite{Feng2017AVE, Wu2019Joint}. The autonomous vehicular edge framework is proposed in \cite{Feng2017AVE} to exploit redundant resources and realize in-vehicle applications. This work also proposes a caching algorithm and adopts the ACO to solve the workload allocation problem. Cache-enabled UAV communications are considered in \cite{Wu2019Joint}, where a rotary-wing UAV is deployed to cache files and serve a number of downlink vehicles. For given solutions of trajectory and content delivery, the PSO is used to solve the binary content placement problem. Simulations show a small optimality gap between the PSO-based scheme and the optimal caching solution. Very recently, the work in \cite{Huang2020Dynamic} considers a caching decision problem for the Internet of vehicles and proposes to deploy a quantum PSO based algorithm at the edge to obtain the caching solutions. 
The interested readers are recommended to follow the survey \cite{dziyauddin2019computation} for more details about vehicular edge computing.

\begin{figure}[t]
	\centering
	\includegraphics[width=0.95\linewidth]{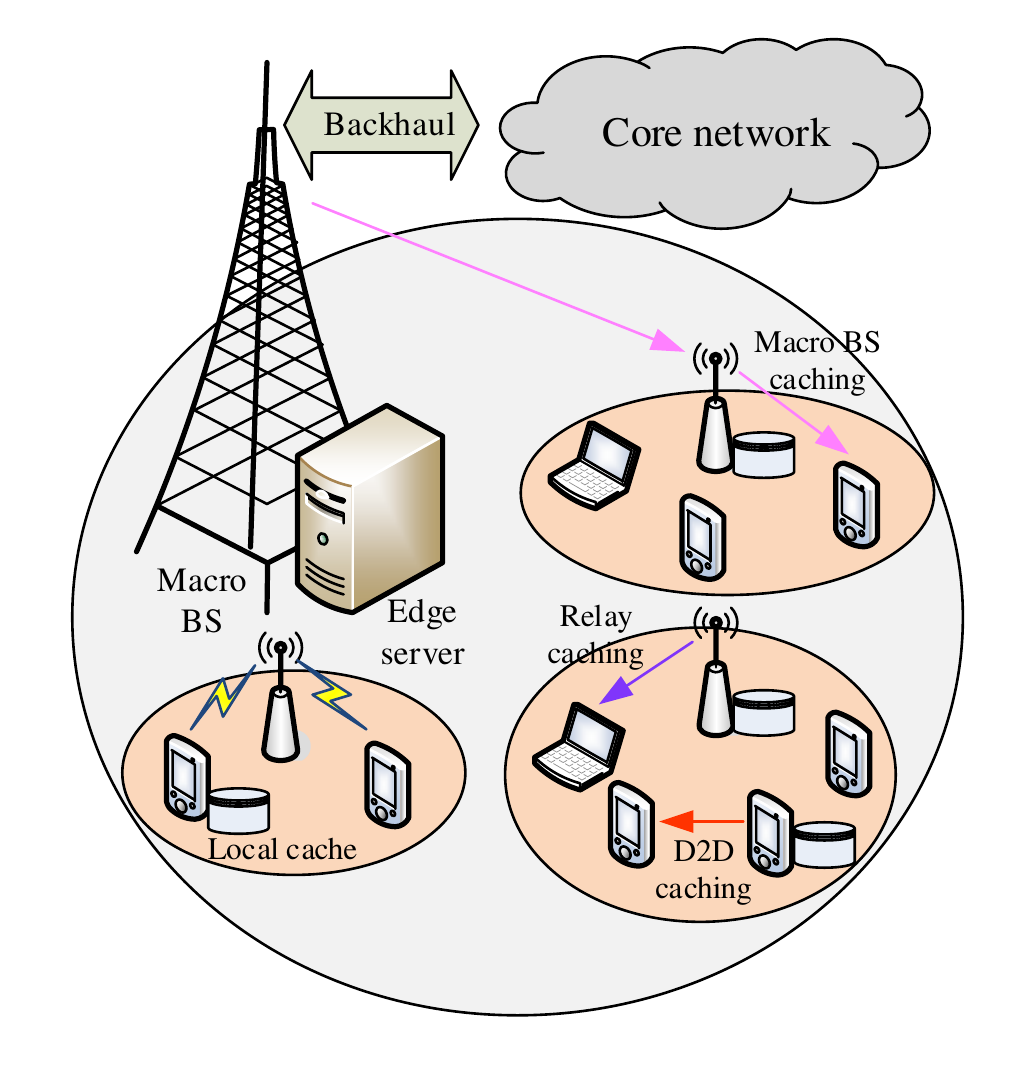}
	\caption{Illustration of a cache-enabled three-tier HetNet, where the content can be cached at the local device, nearby users, relays, and macro BS. Adapted from \cite{Tan2019Heterogeneous}.}
	\label{Fig:Caching_HetNet_D2D}
\end{figure}

HetNet is a key enabler to enhance the performance of cache-based wireless systems. It is shown in \cite{yang2015analysis} that HetNet with caching capabilities (popular contents can be cached at multi-tiers, including macro BS, relay, and nearby users), can achieve a throughput gain of 57.3\% compared with that of the HetNet without caching (i.e., popular contents are cached at the macro BS only). A caching mechanism is investigated in \cite{Tan2019Heterogeneous}, where full-duplex relays are deployed and the mobility patterns of users are modeled to improve the throughput gain. Illustration of a cache-enabled three-tier HetNet is shown in Fig.~\ref{Fig:Caching_HetNet_D2D}. More specifically, there are possibly four cases for a content request, i.e., the requested content is available at 1) the local cache, 2) at least nearby users, 3) nearby relays, and 4) the macro BS. For a given set of cached contents, a PSO-based approach is proposed to solve the joint caching placement problem whose objective is to maximize the network throughput. The main advantages of the PSO are near optimal performance and very low complexity when it is compared with the optimal exhaustive search algorithm. Similar to \cite{Tan2019Heterogeneous}, the work in \cite{ali2020optimal} also exploits the mobility patterns of users to improve the offloading probability, but the use of an adaptive bat algorithm is shown to outperform the PSO-based approach.
In \cite{Zhao2018Analysis}, each mobile user is allowed to have dual connectivity (i.e., one with sub-6 GHz macro BS and the other with mmWave small cells) and the PSO is adopted to maximize the average successful delivery probability. In order to solve the inter-dependencies among the density of HetNets, dynamics of harvested energy, and diverse QoS requirements of users, the work in \cite{Zhang2020ASleeping} considers a caching problem so as to minimize the total energy consumed by the small cell BSs (SBSs) and backhaul links. Furthermore, based on the energy level and traffic offloaded from the macro BS, a binary PSO scheme is studied to decide the sleeping status of SBSs. 

\subsection{Computation Offloading}
\label{SubSec:ComputationOffloading}
Along with the massive connectivity and diverse QoS requirements, \emph{computation offloading} has been regarded as a key solution for many compute-intensive and latency-critical applications and services \cite{pham2020coalitional}. According to the survey \cite{Pham2020Survey_MEC}, besides network optimization, game theory, and AI, many computation offloading (sometimes referred to as task scheduling and task allocation in the literature) problems can be solved effectively by SI techniques with competitive performance and low complexity. Application of bees swarm for computation offloading in fog computing is examined in \cite{bitam2018fog}, which shows significant improvements over the GA and PSO algorithms. The WOA is shown in \cite{pham2020whale} to have very competitive performance compared with the semi-distributed solution that is obtained by the submodular optimization approach. In \cite{mishra2018sustainable}, the service allocation problem for industrial application is optimized by three SI techniques, including the continuous and binary PSO, and bat algorithm. The work in \cite{zhu2019folo} optimizes the task allocation problem to achieve a trade-off between latency and quality in vehicular edge computing systems. More precisely, vehicles can migrate their workloads to other vehicles with computing capabilities and stationary edge servers. The mixed-integer task allocation problem is first solved by the linear programming method, but it has high computational complexity. Therefore, this work also proposes a binary PSO based algorithm to achieve efficient solutions. Using real-world traces to model the mobility of edge servers, the PSO is shown to have lower service latency than the linear programming method when the quality is relaxed, but it is not preferred to use when the quality requirement becomes stricter. The computation offloading problem is also investigated in many other studies \cite{bany2020td, hou2020reliable, Chen2020Delay}. For example, to support latency-sensitive applications on the Internet of vehicles (IoV), the work in \cite{hou2020reliable} considers combining software-defined networking (SDN) and edge computing, and proposes a holistic architecture. Moreover, this work investigates a reliable computation offloading scheme by considering the reprocessing latency. To overcome the difficulties of non-convexity and NP-hardness, the PSO is utilized to design a fault tolerant algorithm. Motivated by the fact that offloading tasks to stationary nodes (e.g., road side units) is highly expensive, the allocation of tasks among vehicles is considered in \cite{Chen2020Delay}. More specifically, a vehicle may offload its computation task to one among multiple idle vehicles and the max-min completion time is optimized. To deal with the challenge that an idle vehicle shall not participate in the tasking processing, a PSO-based algorithm is proposed. Energy-efficient task offloading problems in HetNets are studied in \cite{yang2018distributed, yang2018mobile} with applications of SI techniques. For instance, an algorithm based on artificial fish swarm is developed to minimize the total energy consumption in a three-tier HetNet. The particularity of this work is that the energy consumed by the fronthaul links (between users and relays) and the backhaul links (between relays and edge servers) is considered in the design of fitness function. Recently, different PSO variants are proposed in the literature for task allocation, e.g., improved discrete PSO algorithms in \cite{lin2019time, shao2020task}, a hybrid scheme based on PSO and cat swarm optimization in \cite{rafique2019novel}, and two slow-movement PSO schemes in \cite{zhang2020slow}.

Utilizing IS techniques for computation offloading required by IoT applications has also been received much attention in the last few years. To achieve the load balancing among fog nodes and latency requirements of IoT delay-sensitive applications, two algorithms based on the ACO and binary PSO are investigated in \cite{hussein2020efficient}. A number of simulations are conducted to verify two proposed algorithms, and it is observed that the proposed ACO-based approach performs better than the round-robin and PSO algorithms in terms of average completion time and load balancing. A recent study in \cite{sharma2020fog} considers the problem of task allocation in data deduplication for industrial IoT applications such as air pollution. In particular, a four layer architecture is considered, IoT device layer, fog layer, cloud layer, and service layer. A multi-objective WOA algorithm, including node degree, residual energy, and distance to other devices, is proposed to divide IoT devices to multiple clusters. After that, each cluster head reports hash information (used for checking data duplication) to a fog node, that is selected by another SI approach, namely fast artificial fish swarm optimization. The potentials of SI techniques for IoT offloading applications can also be found in \cite{adhikari2020application, huang2020bilevel, adhikari2019energy}. An accelerated PSO scheme is proposed in \cite{adhikari2020application} for solving the application offloading problem, and the fitness function is designed from multiple parameters such as resource utilization and total (computing and memory) cost. In order to minimize the total energy consumption for latency-sensitive applications, the work in \cite{huang2020bilevel} studies a joint task offloading decision and computation resource allocation problem and proposes a bilevel optimization approach to get the solution. As most offloading decisions are infeasible due to the latency constraints, a prune policy is first proposed to eliminate infeasible solutions. Then, the ACO and monotonic optimization are used to obtain the upper and lower values, respectively, so as to achieve the offloading decisions and to optimize resource allocation. The approach proposed is similar to the branch-and-bound method, where the branching is employed to get the optimal solution for upper-bound optimization. However, it is shown that the bilevel method can approach almost the same performance as the optimal solution, while it is superior to several baseline schemes. Another application of the ACO for IoT computation offloading is studied in \cite{guo2020intelligent}, where one device can migrate its computation tasks to multiple edge servers and the fitness function is designed as a weighted sum of total energy, latency, and computing charge.    

\subsection{Joint Resource Optimization}
\label{SubSec:Joint3C}
The fact is that NGN supports not only communication services as in earlier network generations, but also caching, computing, and control \cite{Pham2020Survey_MEC}. As a result, many studies have been dedicated to addressing various problems about joint resource (caching, computation, and communication) optimization. In this context, the optimization variables can computing resources (e.g., local CPU frequencies and edge computing capabilities), radio resources (e.g., transmit power, subcarrier, and time allocation), and scheduling decisions (e.g., what to cache and where to remotely execute), whereas the objective can be energy consumption, delay minimization, or both \cite{Pham2019Mobile}. The work in \cite{hu2019twin} considers a joint computing, caching, and communication problem in vehicular networks. Due to complex environments and vehicle mobility, a two-time scale is proposed to solve the problem, where the PSO and deep reinforcement learning are used at the large timescale and short timescale, respectively. More specifically, the actions to be optimized at the large timescale are caching placement decisions (i.e., the caching location and size) at the road side units and offloading decisions from the vehicles. In the meanwhile, the actions at the small timescale are the assignment of road side units to each vehicle and the offloading decisions. It is reasonable since a vehicle is now under the coverage of a road side unit, but it shall not be in a future time instance.
Various results show that the proposed twin-timescale algorithm is superior to two baseline schemes: random and equal resource allocation, in terms of cost and the probability of completing the tasks successfully. 
To deal with the mobility and intermittency of edge nodes, the work in \cite{mseddi2019joint} investigates a joint problem of container placement and task scheduling, which aims at maximizing the quantity of satisfied tasks subject to a number of constraints, such as the resource utilization of each fog node should be less than one and the number of requests to all fog nodes of a user should be less than its total load. Two solution approaches are proposed: a greedy approach and a metaheuristic which is designed based on the PSO algorithm. Especially, compared with the Folo algorithm in \cite{zhu2019folo}, this work considers repairing each particle (i.e., solution) if some constraints are violated. Real-world data experiments show that the proposed PSO with reparation achieves performance close to the optimum, especially in the static scenario, whereas it outperforms the Folo in terms of success probability. Applications of the WOA to solve a joint task scheduling and local CPU frequencies are investigated in \cite{peng2019joint}. Interestingly, the WOA can achieve a shorter task completion time compared with the PSO, while both achieve almost the same energy consumption. In a hierarchical edge computing system, the work in \cite{huynh2020efficient} utilizes the PSO to optimize the offloading decisions (i.e., local computing, computing at the small-cell or macro edge servers) and allocates computing resources by a close form solution. Compared with three baseline schemes, including only offloading, offloading without macro edge servers, and offloading only, the PSO-based algorithm can reduce the computation overhead significantly.

\begin{table*}[t]
	\caption{Summary of the state-of-the-art SI techniques for wireless caching and edge computing.}
	\label{Tab:Summary_WC_edgecomputing}
    \resizebox{\textwidth}{!}{	
	\begin{tabular}{|c|c|c|p{12.5cm}|}
		\hline 
		Category & Paper  & Techniques & Highlights  \\ 
		\hline
		\hline
		
		\multirow{5}{*}{\makecell{Wireless \\caching}} & \multirow{1}{*}{\cite{Li2019SociallyAware}}  & \multirow{1}{*}{ACO} & The ACO is introduced to solve the socially-aware caching problem in D2D-enabled radio access network. \\
		
		\cline{2-4} 
		
		& \multirow{2}{*}{\cite{Tan2019Heterogeneous}} & \multirow{2}{*}{PSO} & Four caching scenarios are studied in a full-duplex relying three-tier HetNet, as illustrated in Fig.~\ref{Fig:Caching_HetNet_D2D}, and the PSO is used to optimize the sum throughput of all the users. \\ 
		
		\cline{2-4} 
		
		& \multirow{2}{*}{\cite{Zhang2020ASleeping}} & \multirow{2}{*}{PSO} & A PSO-based algorithm is proposed to minimize the total energy consumption while considering users' QoS, network density, and dynamics of energy harvesting. \\
		
		\hline	
		
		\multirow{6}{*}{\makecell{Computation \\offloading}} & \multirow{2}{*}{\cite{zhu2019folo}}. & \multirow{2}{*}{PSO} & A task allocation problem is formulated to achieve the balance between the completion latency and quality. The PSO is adopted to obtain solutions with competitive performance. \\
		
		\cline{2-4} 
		
		& \multirow{1}{*}{\cite{sharma2020fog}} & \multirow{1}{*}{WOA}  &  A multi-objective WOA approach is introduced to solve the clustering problem for industrial IoT applications. \\
		
		\cline{2-4} 
		
		& \multirow{3}{*}{\cite{huang2020bilevel}} & \multirow{3}{*}{ACO} & A prune policy is first proposed to remove infeasible solutions. Then, the ACO and monotonic optimization are used to in a bilevel framework. The ACO-based algorithm can achieve almost the same performance as the exhaustive search and significantly outperforms baseline schemes.\\ 
		
		\hline
		
		\multirow{7}{*}{\makecell{Joint \\resource \\optimization}} & \multirow{2}{*}{\cite{Tan2019Twin}} & \multirow{2}{*}{PSO} &  A two-timescale method is investigated in vehicular networks, where the PSO and deep reinforcement learning are used at the large timescale and small timescale, respectively. \\
		
		\cline{2-4} 
		
		& \multirow{3}{*}{\cite{peng2019joint, pham2020whale}} & \multirow{3}{*}{\makecell{WOA}} &  The WOA is introduced to solve the joint task allocation and computation resource allocation problem. The WOA-based algorithm is shown to have competitive performances compared with the traditional PSO and the optimal solution. \\ 
		
		\cline{2-4} 
		
		& \multirow{2}{*}{\cite{deng2020incentive}} & \multirow{2}{*}{\makecell{Simulated annealing \\ABC, PSO}} &  Several SI techniques are adopted to solve the joint server selection and computation resource allocation for mobile blockchain services. \\
				
		\hline
	\end{tabular}
    }
\end{table*}

SI techniques have also been considered as a vital tool to solve the optimization problems when different technologies are combined, for example, NOMA, D2D, UAV, and blockchain. The combination of NOMA and MEC with applications of the PSO is considered in \cite{guo2018efficient, seng2018joint}. Firstly, the total utility of all the BSs is optimized by a user association mechanism. This is followed by a resource allocation problem, which aims at maximizing the network energy efficiency and is solved by the PSO to attain an efficient solution at reasonable complexity. The proposed PSO-based algorithm is verified to outperform two existing algorithms: OMA and a mechanism that uses the minimum-distance as the design metric. The work in \cite{zhu2018cooperative} minimizes the energy consumption by jointly optimizing radio and computing resources under constrains on latency and spectrum restriction. In particular, a UAV shall partially offload the task to one among multiple terrestrial edge servers, which may cooperate with the other servers to process the received workload. The PSO and simulated annealing are combined to obtain the solution for a computation resource allocation problem, which would avoid the trap of falling into local solution by the original PSO. SI techniques have also found potentials for blockchain-based edge computing systems. It is elaborated in \cite{deng2020incentive} that blockchain is a vital technology to overcome the main challenges of e-commerce services, including trust issues, trading security, data privacy, and logistics. To enable mobile blockchain for e-commerce services, a joint server selection and computing resource allocation problem is investigated, which is then solved by several metaheuristic approaches, including GA, simulated annealing algorithm, ABC, and PSO. Another work on blockchain for cloud manufacturing with applications of the PSO can be found in \cite{yu2019blockchain}.

\subsection{Summary}
\label{SubSec:WirelessCaching_Summary}
The applications of SI for wireless caching and edge computing problems are reviewed in this section. More specifically, we review three main problems in edge computing systems, including wireless caching, computation offloading, and joint resource optimization. Since edge systems comprise of many optimization variables along with environment dynamics and mobility, typically problems are very challenging to be solved to get the globally optimal solution. SI techniques have great potentials in solving these problems. Similar to the earlier section, we observe that the ACO and PSO optimizers are widely used in the literature thanks to their high performance and originality (i.e., ACO is originally developed to solve discrete problems). However, recent approaches still find their applications and advances in improving the network performance. As an example, the WOA is shown in \cite{pham2020whale, peng2019joint} to have competitive performance over the traditional PSO scheme. The review in this section opens many opportunities to use SI techniques for optimizing edge computing when it is combined with new NGN technologies such as THz communications and reconfigurable intelligent surface. To summarize this section, representative reviewed references, used techniques, and highlights are presented in Table~\ref{Tab:Summary_WC_edgecomputing}.

\section{Network Security}
\label{Sec:DataPrivacy}
Security is always one of the key concerns for any network generation, but this issue has become a task due to the changes and complexities in underlying network and the emergence of many technologies and services. According to \cite{s1}, there are three major features of 5G security, including 5G security requirements inherited from threats in previous network generations, new security challenges caused by massive device connectivity and heterogeneous wireless communication, and new security issues from new technologies like SDN, MEC, network function virtualization (NFV), cloud computing. In this section, we discuss the applications of swarm intelligence in dealing with security problems in 5G networks.

\subsection{Access Control}
Access control mechanisms aim to prevent malicious users and attacks from accessing resources, e.g., computing or storage resources. The use of access control approaches such as attribute-based access control is important for ensuring the flexible network management against security attacks, by allowing only legitimate users and devices to use the network resource while blocking any external threats for security improvement. SI is able to provide viable solutions to achieve efficient access control for 5G networks. For example, an extended attribute-based access control method is proposed in \cite{s3} to enhance the confidentiality and integrity of SDN-empowered 5G networks. For improving the security of the access data flow in the SDN switch, a PSO algorithm is used to obtain the optimal path of multi-level security and multi-switching nodes where SDN switches are considered as environmental attributes. The simulation results confirm high security capability achieved by PSO with a low network response delay. Another effort in using PSO algorithms for access control in SDN is in \cite{s4}. More specifically, an enhanced PSO-based routing protocol with a cuckoo search is proposed to support SDN controllers in route selection with respect to the available links, latency, and controller load balancing. The key purpose of the proposed scheme is to realize the optimal routing for improving the QoS while minimizing the malicious user access possibility with efficient threat prevention. 

Moreover, SI-based algorithms can provide access control for cloud-based 5G networks. In fact, how to implement intelligent user access control in the cloud computing environments is highly important in the context of complex user patterns (e.g., cloud data storage, cloud data computing, and cloud data flow tracking). SI has emerged as a promising technique to enable smart data access analysis for security guarantees \cite{s5}, by supporting cryptography algorithms in deducing optimum keys based on statistical probability for cryptanalysis that is necessary for network transmission monitoring.  In addition to that, SI is also applied to cloud scheduling tasks, including access control, for user data services (e.g., data offloading) \cite{s6}. In this case, an ACO-enabled scheduling algorithm is derived to solve data-intensive scheduling problems, i.e., task level scheduling and access reliability levels, with respect to overall throughput, running cost, scheduling deadline. The simulation results from using the ACO scheme with cloud datasets in virtual machines confirm a better security level compared to the traditional baselines. Further, SI-empowered algorithms such as PSO can also build algorithms for providing data access reliability in cloud storage \cite{s7}, by optimizing the data transfer trust and storage stability under cloud storage capability and storage cost. 
\subsection{Data Privacy}
In 5G networks, network operators and mobile users interconnect and form business models where a large amount of information can be exchanged over the networks. The ubiquitous information sharing among data users over the mobile networks may raise many critical privacy concerns such as data leakages and user information attacks. Therefore, 5G networks must consider an end-to-end data protection approach to provide high degrees of user privacy where SI can provide highly efficient solutions. Indeed, SI-based methods such as PSO can build security-preserved network models for 5G IoT networks \cite{s10} that can be divided into a multi-layer clustering system using decentralized blockchain technologies. At each cluster, a local data privacy mechanism is designed to perform self-clustering for IoT that optimizes the network lifetime with the data integrity enhancement. PSO is also applied to solve different privacy in data clustering with a particle subswarms collaborative clustering scheme \cite{s12}. The aim is to group particles to enhance the collaboration between subswarms in a fashion the data sets are processed separately and only coefficient information is exchanged. This would solve the data clustering problem without compromising data privacy requirements. To realize fractal intelligent data privacy preservation in social IoT networks, PSO has been considered in \cite{s13} thanks to its high optimization ability by simulating bird predation behavior. A social data privacy protection model is formulated based on the security situation prediction scheme, and a PSO algorithm is adopted to solve the formulated privacy problem, showing a significant improvement in the privacy protection accuracies over the traditional schemes, such as semisupervised learning methods and matrix factorization schemes. 

\begin{figure*}
	\centering
	\includegraphics[width=0.625\linewidth]{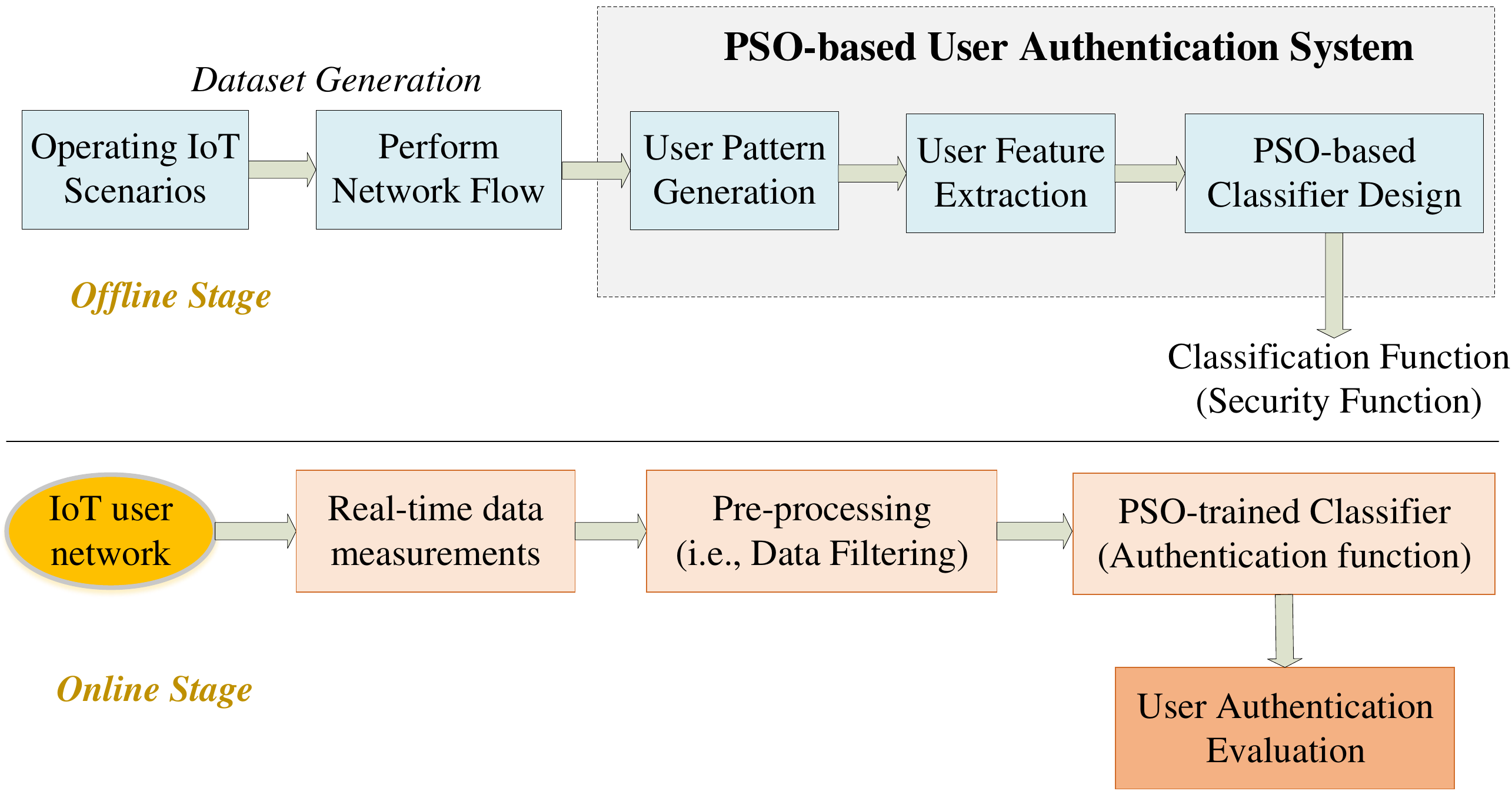}
	\caption{An illustration of a PSO-based model for user network authentication. }
	\label{Fig:SI_security}
\end{figure*}

In 5G technologies, data privacy protection is an urgent problem. The work in \cite{s14} shows the first attempt to use SI techniques in order to preserve data privacy in SDN-based 5G networks. The current SDN routing schemes often focus on data packet transfer without privacy concerns, e.g., information leakage, in the communication channels among SDN controllers and users. SI-based algorithms such as ACO can arrive to build distributed routing optimization models by taking privacy exposure and risk definition into consideration. Each controller selects some different features such as data packet types, network topology, and data routing history, which are classified by K-mean learning. Then, these classification outputs are used as the parameters of the ACO-based real-time routing scheme, aiming to find the optimal route with minimum privacy risks. The application of SI to deal with privacy issues in cloud-based 5G networks is also considered in \cite{s15}. In cloud computing, the real-time data transmission and the complex data usage patterns make it challenging to protect data privacy against external attacks and cloud third party. Therefore, a new dragon PSO algorithm is suggested to establish a k-anonymization criteria to derive the fitness function that represents the privacy and utility levels of cloud migration services. The experiments using an Adult database verify the efficiency of the privacy-preserved model against information loss risks \cite{s15}.  Moreover, the work in \cite{s16} presents a data privacy-preserved PSO algorithm for cloud computing. Especially, the traditional cloud is replaced by a cloud microservice that is deployed at the mobile device instead of remote cloud servers. In this regard, a quality of experience-aware model is developed with a focus on minimization of application response time with privacy awareness.

\subsection{Network Authentication}

Network authentication mechanisms are used to authenticate the identity of an entity, e.g., data requestors, data access users, in order to detect and prevent malicious network attacks. SI can provide new intelligent network authentication solutions for mobile networks towards 5G. For any authentication modes in wireless networks, the evaluation of authentication depends on two main classes: authorized and unauthorized. SI can be used to build intelligent user authentication systems in wireless networks, as illustrated in Fig.~\ref{Fig:SI_security}. Here, SI algorithms such as PSO can perform user classification in off-line and on-line stages based on user patterns, e.g., access frequency, data usage behaviors, in order to determine whether this user is a legitimate network entity or not. Recently, some research efforts have been made to apply SI techniques to improve network authentication. For example, the study in \cite{s18} builds a user authentication system to capture the user patterns in minimal time including two phases: the enrolment phase for data capturing and pattern learning, and the verification phase for feature extraction and comparison with the biometric pattern. In particular, the feature selection is performed by a PSO algorithm associated with support vector machine and Naïve Bayesian in keystrokes dynamic authentication environments for fast user feature detection. To perform security protocols in mobile ad hoc networks (MANET), an ACO-based approach is proposed in \cite{s19} with the objective of securing communications and authenticating users in MANET in the presence of attacks while ensuring low network latency. The ACO algorithm is able to learn the optimal routing path with data reliability (e.g., authenticated data access) and QoS preservation. Similarly, the work in \cite{s20} presents a network authentication approach enabled by a trusted QoS protocol in MANET. An intelligent PSO algorithm is derived to learn the packet delivering procedure for identifying and isolating unauthorized nodes using the user routing information. If the trust weight is below a predefined threshold, the target node is regarded as a malicious node; otherwise, its authentication is validated for communication establishment. Meanwhile, a joint framework using ACO and elliptic curve cryptography is considered in \cite{s21} for vehicular ad hoc network (VANET). The focus is on analyzing some primary issues in VANET, such as security improvement including data validation and authorization services, and randomness minimization of existing traffic management systems. For security provision, a mutual authentication scheme is integrated with learning using radio frequency identification to detect intelligently any false vehicular communication messages (i.e. wrong traffic messages) in dynamic vehicular settings. 

In 5G cognitive radio networks, how to enable secure and trusted communication between transmitters and receivers is highly important. Some learning-based solutions such as PSO may be very useful in the authentication classification and attack verification \cite{s22}. In this work, a mutual authentication mechanism is designed where the user identity and real identity are produced by both the transmitter and receiver for authentication validation enabled by a PSO-based access feature selection. The simulation results confirm a high user authentication level (i.e., detection accuracy) with improved system throughput and enhanced packet delivery ratio, compared to the traditional schemes. Enabled by SI that forms a closed-loop system like swarms in humans when the groups can perform better in analyzing and achieving the consensus, an SI-inspired intelligence approach is also proposed in \cite{s23} to deal with cognitive radio authentication. Three use cases are employed to evaluate with vehicular fog, including measurable security, privacy, and dependability of autonomous vehicles, showing a better security level and more reliable authentication in comparison with non-intelligence schemes. SI has been also adopted in \cite{s24} to build a self-defensive multi-layered cloud computing platform that is promising to develop self-aware 5G systems where entities such as users and service providers can work together for cooperation in network operations including authentication, based on the concept of SI. 

\begin{table*}
	\centering
	\caption{Summary of the state-of-the-art SI techniques for network security. }
	\label{Tab:Summary_Table_Security}
	\begin{tabular}{|P{1.8cm}|c|c|p{12.5cm}|}
		\hline
		\centering {Category}& 	
		{Paper} &	
		\centering {Techniques}&	
		{Highlights}	
		\\
		\hline
		\hline
		\multirow{8}{1.5cm}{Access control} & 
		\multirow{2}{*}{\cite{s3}} & \multirow{2}{*}{PSO} & An attribute-based access control method is proposed using PSO to enhance the confidentiality and integrity of SDN-empowered 5G networks. 
		\\ \cline{2-4}&
		\multirow{2}{*}{\cite{s4}} & \multirow{2}{*}{PSO} & A method using PSO algorithms is introduced for access control in SDN, aiming to realize optimal routing for improving (QoS) while minimizing the malicious user access possibility with efficient threat prevention.
		\\ \cline{2-4}&
		\multirow{2}{*}{\cite{s6}} & \multirow{2}{*}{ACO} & An ACO-enabled scheduling algorithm is derived to solve data-intensive scheduling problems, i.e., access control, with respect to overall throughput, running cost, scheduling deadline. 
		\\ \cline{2-4}&
		\cite{s7}	&PSO&	A PSO-empowered algorithm is designed for providing data access reliability in cloud storage. \\
		\cline{2-4}	
		\hline
		
		\multirow{4}{1.5cm}{Data privacy} & 
		\cite{s10} &	PSO	&A PSO algorithm is built for security-preserved network models in 5G IoT networks.
		\\ \cline{2-4}&
		\cite{s13}&	PSO	&A PSO-based scheme is provided to realize fractal intelligent data privacy preservation in social IoT networks. 
		\\ \cline{2-4}&
		\cite{s14}&	ACO&	An ACO-based framework is designed to preserve data privacy in SDN-based 5G networks. 
		\\ \cline{2-4}&
		 \multirow{2}{*}{\cite{s15}} & \multirow{2}{*}{PSO} & A new dragon PSO algorithm is suggested to establish a k- anonymization criteria to derive the fitness function that represents the privacy and utility levels of cloud migration services.
		\\ \cline{2-4}	
		\hline
		
			\multirow{8}{1.5cm}{Network authentication} & 
		\multirow{2}{*}{\cite{s19}} & \multirow{2}{*}{ACO} & An ACO-based approach is proposed with the objective of securing communications and authenticating users in MANET in the presence of attacks while ensuring low network latency. 
		\\ \cline{2-4}&
		\multirow{2}{*}{\cite{s21}} & \multirow{2}{*}{ACO} & A joint framework using ACO and elliptic curve cryptography is considered for vehicular ad hoc network (VANET) for security improvement including data validation and authorization services.
		\\ \cline{2-4}&
		\multirow{2}{*}{\cite{s22}} & \multirow{2}{*}{PSO} & A learning-based solutions such as PSO is introduced for the authentication classification and attack verification in wireless networks. 
		\\ \cline{2-4}&
		\multirow{2}{*}{\cite{s23}} & \multirow{2}{*}{SI} & An SI-inspired intelligence approach is proposed to deal with cognitive radio authentication. This scheme can provide better security levels and more reliable authentication. 
		\\ \cline{2-4}	
		
		\hline
		\multirow{8}{1.5cm}{Attack detection} & 
		\multirow{2}{*}{\cite{s28}} & \multirow{2}{*}{ABC} & An ABC-based optimization classifier is built to perform classification of the cloud data packets for intrusion identification that is potential to solve DoS attacks. 
		\\ \cline{2-4}&
		\multirow{2}{*}{\cite{s29}} & \multirow{2}{*}{SI} & A dolphin SI is investigated where the biological features associated with SI can be used to optimize the detection and accuracy of intrusion identification. 
		\\ \cline{2-4}&
		\multirow{2}{*}{\cite{s30}} & \multirow{2}{*}{WOA} & A WOA scheme is proposed for intrusion detection in wireless networks that can achieve better classification accuracy with using fewer feature subsets.
		\\ \cline{2-4}&
		\multirow{1}{*}{\cite{s31}}	&\multirow{1}{*}{PSO} & A binary PSO algorithm is designed  to identify unauthorized users and detect data attacks in SDN.
		\\ \cline{2-4}	
		\hline
	\end{tabular}
\end{table*}

\subsection{Attack Detection}
Attack detection aims to identify and prevent any attacks and threats to wireless networks. In the context of 5G where attackers become much more sophisticated, intelligent attack detection solutions are of paramount importance for 5G security. It is believed that SI can provide feasible solutions for efficient attack detection in 5G.  Indeed, SI can build security decision making models to fight against malicious threats and intrusions based on security postures and intelligence information using full-text search, mathematical optimization, logical reasoning, and probability analytic. For a typical use case, the work in \cite{s28} tries to adopt SI for intrusion detection in cloud computing. An ABC-based optimization classifier is built to perform classification of the cloud data packets for intrusion identification that is potential to solve security issues such as denial-of-service attack (DoS attack), replay attack, and flooding attacks in cloud computing. For intrusion detection systems (IDS) in VANET, dolphin swarm intelligence has been investigated in \cite{s29} where the biological features associated with SI can be used to optimize the detection and accuracy of IDS. Also, the dolphin swarm behaviors such as hunting and spraying are demonstrated the feasibility for detection and prevention of malicious nodes, e.g., malicious vehicles, and data attacks from the VANET. The SI-enhanced IDS is also considered in \cite{s30} for wireless networks. Different from \cite{s29}, this study relies on a WOA scheme that can achieve better classification accuracy by using fewer feature subsets. The simulation from KDD CUP 99 dataset shows its efficiency in improving intrusion detection accuracy, compared to its counterparts such as the PSO algorithm. 

SI has been applied to attack detection tasks in SDN-based 5G networks. For instance, a binary PSO algorithm is designed in \cite{s31} to identify unauthorized users and detect data attacks in SDN. First, a security rule is defined by using an ML-based random forest algorithm and prediction results obtained from the training procedure with the KDDTest21FS dataset. Then, PSO is applied to feature selection that is necessary for user classification. The proposed scheme not only achieves high attack identification accuracy but also brings a lower detection sensitivity. Another research effort for attack detection in SDN is in \cite{s32}. Here, the authors pay attention to attack detection in SDN controllers, by developing a bio-inspired algorithm as a data flow management function for user access monitoring. In terms of Distributed Denial of Service (DDoS) attack detection in SDN, the work in \cite{s33} suggests an optimized PSO learning approach. In this case, how to differentiate the DDoS attack traffic from the genuine traffic is a challenge. Motivated by this, a PSO-based classification model is derived to classify the network traffic patterns as normal or attack traffic. The real-time implementation of the proposed system model indicates the practicality of the proposed SI algorithm for handling the attacks and minimizing the data loss risk for legitimate users.

\subsection{Summary}
This section reviews the applications of swarm intelligence for problems related to network security. The most popular SI technique used in 5G security is PSO that can provide flexible and intelligent solutions for security in 5G technology platforms such as cloud, MEC, SDN, and NFV. Reviewing the state-of-the-art literature in the field, we discuss the roles of SI in four key 5G security services, including access control, data privacy, network authentication, and attack detection. In fact, SI is able to improve the security of the access data flow in 5G networks, e.g., SDN systems, where PSO-based algorithms are useful to provide multi-level access control mechanisms, aiming at minimizing the malicious user access possibility with efficient threat prevention \cite{s3,s4}. Enabled by the high optimization ability by simulating bird predation behavior, SI-based methods such as PSO can build security-preserved network models for 5G IoT networks to promote universal data exchange without compromising data privacy requirements \cite{s12}. Recently, some data privacy-preserved SI algorithms are also considered for SDN and cloud computing-based 5G networks \cite{s16} by forming service migration functions with privacy awareness. Other SI techniques such as ACO \cite{s19} also facilitate the 5G network authentication with bio-inspired algorithms embedded at the network controllers, e.g., SDN controllers, for securing communications and authenticating users with QoS preservation. In addition to that, the potential of SI is also reflected in the attack detection tasks thanks to its learning and classification capability \cite{s29}, such as in the intrusion detection systems of 5G IoT networks. The discussion in this section highlights the advantages of SI techniques in solving security issues in 5G and opens new opportunities for emerging SI-inspired security applications. To summary this section, the reviewed literature, used techniques, and highlights are summarized in Table~\ref{Tab:Summary_Table_Security}.

\section{Miscellaneous Issues}
\label{Sec:Miscellaneous_Issues}
In this section, we survey the use of SI in many other applications in 5G networks, including antenna design for 5G communications, UAV placement and path planning, clustering and routing in IoT networks, smart city applications, and energy management in smart grid.

\subsection{Antenna Design}
\label{SubSec:AntennaDesign}
Recently, SI techniques have been applied to facilitate antenna design. For instance, the work in \cite{m1} presents a solution for tri-band multi-polarized adaptive array antenna in 5G base station. To optimize the antenna parameters and synthesizing multi-beam patterns, a PSO algorithm is adopted with an enhanced version of a hybrid gravitational search scheme for improving the global search ability and speed up the overall algorithm convergence. The simulation results reveal that the improved PSO scheme is able to increase the antenna gain and bring a better coverage efficiency for phased array antenna at different frequency bands in mm-wave communications. Another SI solution is also introduced in \cite{m2} for microstrip antenna design that can tackle the problem of mm-wave antenna impedance mismatch in 5G communication systems. An ACO-based algorithm is employed that uses the pheromone guidance mechanism to achieve the global optimal solution by optimizing patch parameters. By simulating the model at 28GHz centre frequency, the ACO scheme can effectively achieve impedance matching and improve the return loss characteristics. The SI-based ABC approach also proves its usefulness in antenna design for wireless communication systems \cite{m3}. Three steps in the ABC algorithm are exploited specific to antenna design, including bee stage, onlooker bee step, and scout bee step. The first two steps aim to explore the potential optimal solutions, while the last step ensures the convergence of the algorithm. Especially, a similarity induced search method is integrated to adaptively adjust search step size, which is promising to achieve scalability for the antenna design problem in terms of array elements. 

The potential of SI is also verified in \cite{m4} where the design of rectangular patch micro strip antenna is considered. The radiation pattern of strip antenna is derived using a PSO algorithm within a short simulation period. In \cite{m5}, a metaheuristic PSO model is developed for patch antenna design, by optimizing the operating frequency, the reflection parameter, and the impedance of the antenna. Compared to the baselines such as the firefly scheme and directional bat scheme, the PSO scheme can achieve the best performance in terms of length, width, and insertion distance of the antenna at a fast convergence rate. The authors in \cite{m6} apply SI in designing an MIMO antenna by using the optimal dimensions of the antenna elements. A bow tie antenna as the element of the MIMO is selected with the center frequency at 28GHz that is then optimized by a bio-inspired algorithm called Salp SI for improving the wideband operations in the targeted frequency band. 

\subsection{UAV Placement and Path Planning}
\label{SubSec:UAV_Placement}
In this section, we discuss the benefits of SI approaches in the placement and path planning design of UAV in NGN. 

\begin{figure*}
	\centering
	\includegraphics[width=0.825\linewidth]{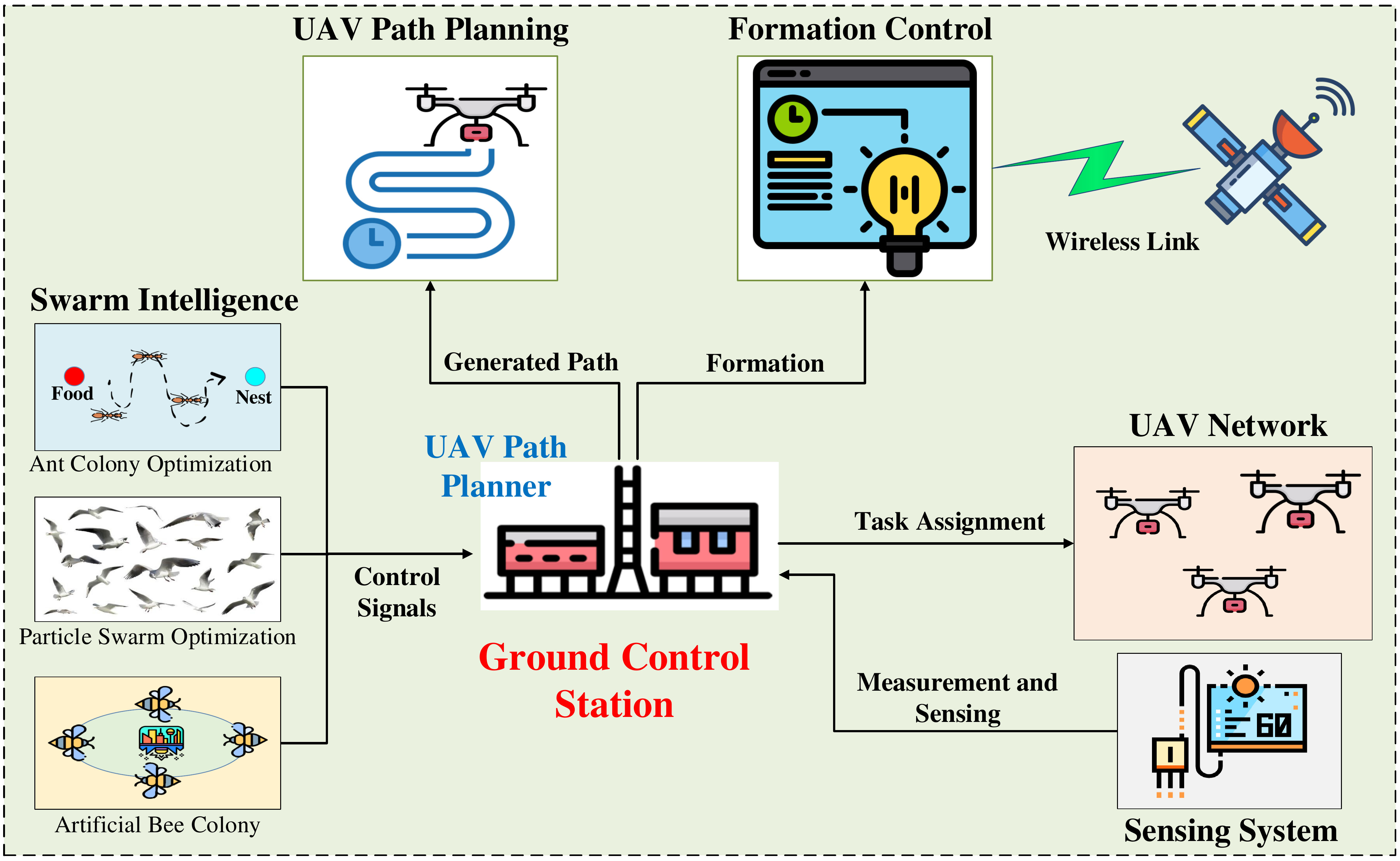}
	\caption{The UAV path planning system with swarm intelligence optimization, adapted from \cite{m11}. }
	\label{Fig:SI_UAV}
\end{figure*}

\subsubsection{\textcolor{black}{UAV Placement}}
By relying on the underlying concept of self-organized behavior in nature, SI algorithms have proved their great potential in improving UAV-related communication designs, such as UAV placement.  The work in \cite{m7} presents a 3D UAV placement framework for wireless coverage of the indoor environments in high-rise buildings under disaster situations, i.e., earthquakes. To optimize the transmit energy consumed to cover the building, a PSO-based algorithm is designed that can find exactly the location of UAV under the random distributions of users. The 3D UAV placement for indoor users in small coverage areas is also considered in \cite{m8} whose model can be used in wireless convergence for massively crowded events. Here, the UAV altitude placement is formulated with the objective of minimizing the transmission power and achieving the optimal altitude for both single UAV and multiple UAVs cases. Another UAV placement solution for massively crowded events is studied in \cite{m9}. UAVs can be used to provide wireless coverage for indoor and outdoor users by using the ground base stations. To specify the optimal 3D UAV placement, both Air-to-Ground and Outdoor-to-Indoor path loss models are employed to optimize the data rates of indoor and outdoor users under a UAV transmit power budget. Meanwhile, the research in \cite{m10} focuses on the 3D UAV placement problem with QoS awareness in ad hoc wireless networks. The number of aerial base stations and transmit power of each station are jointly formulated and then solved by a PSO optimization algorithm in an iterative manner. The simulation results show a significant improvement of performance gain of the joint scheme that can provide an energy-efficient solution for 3D UAV placement of base stations in an ad hoc network. 

\subsubsection{\textcolor{black}{UAV Path Planning}}
A bio-inspired optimization learning environment can be built to simulate the UAV path-planning models \cite{m11} as shown in Fig.~\ref{Fig:SI_UAV}. Here, SI approaches such as ACO, PSO, and ABC are exploited to perform a student learning process for supporting the UAV path planner located at the ground control station. In this regard, by combining the control signals generated by SI algorithms and measurement and sensing parameters obtained from sensors, the UAV path planner can make an intelligent UAV path planning with optimal routing for transmit power minimization. Based on the statistical analytic results, these SI-inspired algorithms show superior advantages in terms of strong robustness in comparison with directed-search approaches with simplicity of implementation and strong adaptability in different UAV path planning scenarios. Meanwhile, an improved ACO-based UAV path planning architecture is proposed in \cite{m12} with obstacles taken into account in a controlled area with several radars. To build a feasible path for UAVs, a travelling salesman problem is formulated via a tour construction of an ant that mimics the salesman routing in cities. To this end, a multi-colony ACO algorithm is built to solve the formulated path planning problem, showing a better performance than the classical ACO scheme. Moreover, the path planning issue for UAV swarms is also solved in \cite{m13} where each UAV is regarded as an intelligent agent which performs searching its targets by updating the position and velocity in reconnaissance contexts. A distributed PSO model is taken to generate the optimal paths for UAVs for fast convergence and minimal latency for target detection. The work in \cite{m14} also employs a set-based PSO algorithm with adaptive weights to implement an optimal path planning scheme for a UAV surveillance system, aiming to minimize the energy consumption and improve the disturbance rejection response of UAVs. 

\subsection{Clustering and Routing in IoT Networks}
\label{SubSec:SmartCity_IoT_Clustering}
The clustering and routing problems have become mandatory for a variety of wireless IoT networks, for example, mobile/flying ad hoc networks,  wireless sensor networks, and complex communication networks. SI has the potential to provide flexible and intelligent clustering and routing solutions for IoT networks. 
\subsubsection{\textcolor{black}{Clustering in IoT Networks}}
Recently, some research efforts have been devoted to use SI techniques to solve clustering issues in wireless IoT networks. For example, the work in \cite{m15} suggests an SI-based clustering framework for wireless IoT sensor networks. The focus is on the energy efficiency improvement and energy usage balance through a fitness function that is optimized by a chicken swarm optimization algorithm. Meanwhile, a hierarchical clustering approach is proposed to implement the data clustering of the sanitized dataset in a fashion the confidential IoT data is kept while allowing to discover useful information \cite{m16}. In this regard, a multi-objective PSO algorithm is adopted as a sensitization method to hide confidential information. Furthermore, to acquire a better positioning accuracy of the clustering tasks in location identification within MIMO sensor networks, a PSO algorithm is derived in \cite{m17} to optimize the cluster head in terms of reduction of calculation complexity and clustering decision convergence  improvement. The authors in \cite{m18} present a mobile sensing cluster using SI to create multiple swarms in mobile devices. This solution is promising to solve the turnaround searching latency issue caused by the single swarm configuration and creates new ways for dynamic multiple swarming for multiple sensing events in a given time. 

\subsubsection{\textcolor{black}{Routing in IoT Networks}}
The work in \cite{m19} implements an optimized routing model for Internet of Medical Things (IoMT) where the main objective is laid towards communication performances in terms of packet delivery ratio and energy constraints. In particular, an ABC algorithm is employed to perform the selection of optimal routing between the source and destination IoMT nodes in order to improve both packet delivery ration and energy usage. In terms of routing in vehicular networks, the authors in \cite{m20} apply an ACO scheme to enable a distributed intelligent traffic system for connected vehicles. Each vehicle can act as an intelligent agent to make routing decisions for minimizing waiting time in automated driving tasks. Besides, in wireless IoT sensor networks, how to perform flexible routing for traffic congestion control is highly important to enhance the QoS provisions regarding network lifetime and data packet drop ratio. The PSO machine learning algorithm can be employed in these scenarios to optimize a fitness function by implementing a local searching using swarm particles and node discovery \cite{m21}. The simulation results confirm a better routing performance than that in the ACO and ABC approaches. PSO is also applied in \cite{m22} for energy-efficient routing in IoT networks. Thanks to its adaptive transmission optimization ability, PSO algorithms are able to perform bio inspired swarm intelligence for cluster head selection with energy awareness. This would facilitate the network scheduling to find the best routing for IoT nodes, aiming to reduce the packet transmit latency and improve energy efficiency. 

\begin{figure*}
	\centering
	\includegraphics[width=0.725\linewidth]{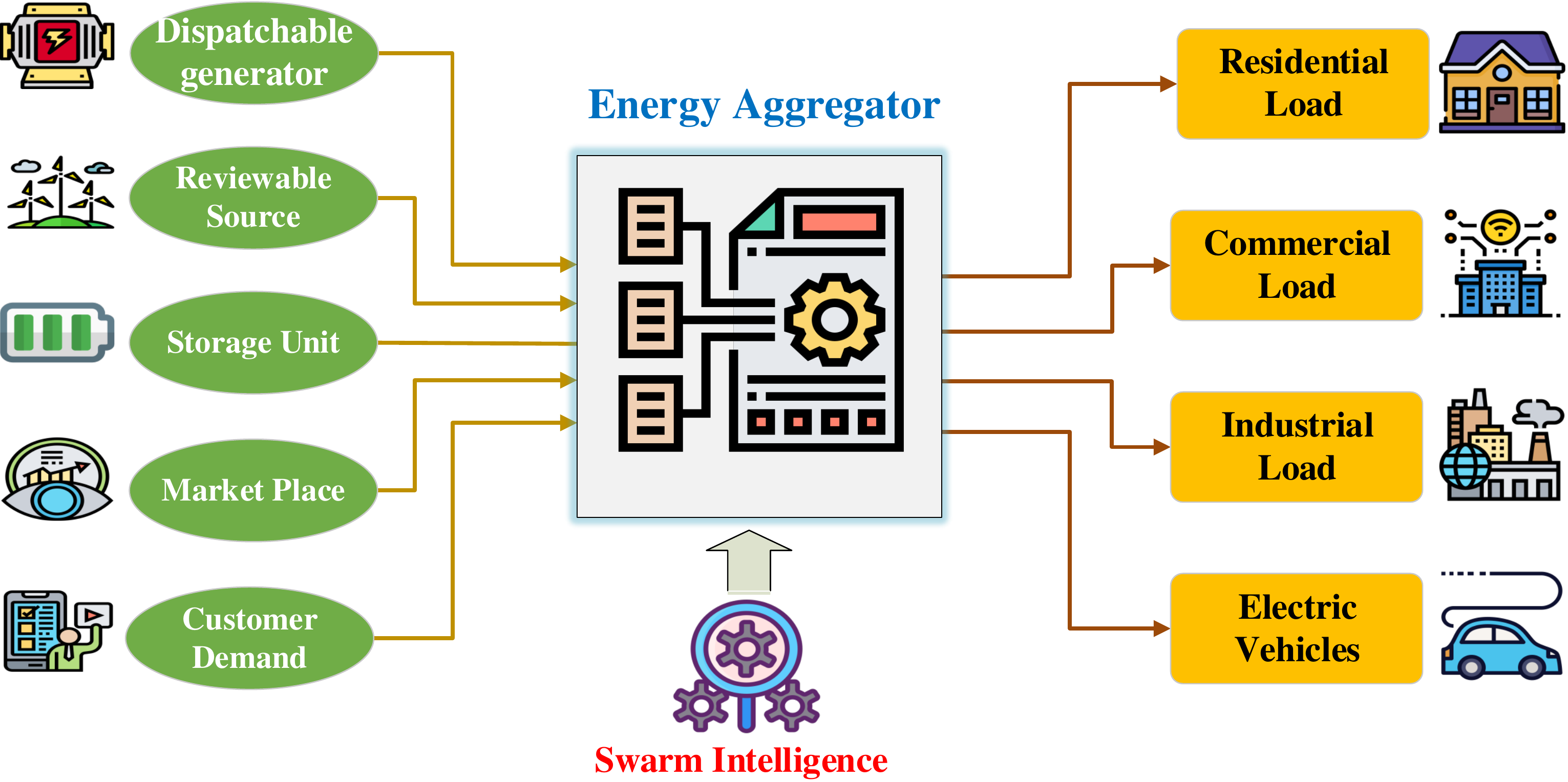}
	\caption{Swarm intelligence optimization for energy resource management in smart grids \cite{m33}. }
	\label{Fig:SI_SmartGrid}
\end{figure*}

\subsection{Smart City Applications}
\label{SubSec:SmartCity_Application}
SI techniques have been also applied to enable smart city applications. The work in \cite{m24} presents a first attempt to use PSO optimization algorithms for a smart city in the context of smart communities. Here, both the power supply and power consumption models are taken into account where energy costs, actual electric power load at high load hours, and the amount of carbon dioxide emission are jointly optimized. The model is investigated at Toyama city, Japan, showing a performance enhancement in terms of lower energy consumption and emission, compared to other schemes like differential evolution.  Another work pays attention to using SI for privacy guarantee in smart power systems in smart cities \cite{m26}. In practice, the energy usage patterns of customers can be disclosed in the smart grid operations, i.e., energy consumption calculation process. Then, PSO can arrive as a solution to create group consumption patterns that can hide individual consumer characteristics for privacy preservation. 

SI has been used to support other smart city services such as vehicular networks. As an example, the work in \cite{m27} provides an SI-enabled solution for smart mobility and autonomous driving in smart cities with edge, fog, and cloud computing. Each vehicle can act as an intelligent agent in the SI algorithm that can estimate the traffic conditions and vehicle diversity for driving control and traffic routing. SI has been adopted in \cite{m28} to predict the traffic flow in smart cities. A PSO algorithm is essential to select the learning parameters in the support vector machine-based prediction model. In addition, to support vehicular clustering process, a GWO technique is proposed in \cite{m29} where the number of vehicular cluster is optimized with respect to grid size, load balance factor, vehicle speed, and movement range. Since the algorithm is inhered from the social nature of gray wolves, the proposed scheme is able to explore better the searching space for finding the more concise optimal solution that is promising to achieve high vehicular routing performances in terms of reduced traffic delay and better network resource usage. 

\subsection{Energy Management in Smart Gird}
\label{SubSec:EnergyManagement_SmartGrid}
In addition to the above use case domains, SI has been also useful in energy management for smart grid. For instance, to facilitate the cost-efficient power management in multi-source renewable energy microgrids, a PSO algorithm is used in \cite{m30}. The key objective is to minimize the daily power cost of renewable microgrids, by finding the maximal energy mixing rate with PSO under energy balance and anti-islanding conditions. The implementation results using PSO confirm a performance enhancement with a better energy efficiency along with the flexible adjustment of mixing energy rate. The use of SI for energy management is also investigated in \cite{m31}. Here, a binary PSO algorithm is developed to implement the home energy management scheduling by following the energy consumption patterns of customers in a fashion the total power cost is minimized while satisfying user comfort. A similar SI-based energy management model is studied in \cite{m32} for energy storage systems in microgrids. The main goal is focused on improving the operational efficiency that can be achieved by using PSO with respect to power constraints and energy flow constraints. 

\begin{table*}[t]
	\centering
	\caption{Summary of the state-of-the-art SI techniques for other applications in 5G networks. }
	\label{Tab:Summary_Table_Miscellaneous}
	\begin{tabular}{|P{1.8cm}|c|c|p{12.5cm}|}
		\hline
		\centering {Category} & {Paper} & \centering {Techniques} &	{Highlights} \\
		\hline
		\hline
		\multirow{6}{1.5cm}{Antenna design} & 
		\cite{m1} & PSO & A PSO algorithm is presented for designing tri-band multi-polarized adaptive array antenna in 5G base station. 
		\\ \cline{2-4}&
		\multirow{2}{*}{\cite{m2}} & \multirow{2}{*}{ACO} & An ACO algorithm is proposed for microstrip antenna design that can tackle the problem of mm-wave antenna impedance mismatch in 5G communication systems. 
		\\ \cline{2-4}&
		\multirow{2}{*}{\cite{m3}} & \multirow{2}{*}{ABC} & An SI-based ABC approach is designed for antenna design in wireless communication systems, aiming to achieve high scalability for antenna design problem in terms of array elements.
		\\ \cline{2-4}&
		\multirow{2}{*}{\cite{m5}} & \multirow{2}{*}{PSO} & A metaheuristic PSO model is developed for patch antenna design, by optimizing the operating frequency, the reflection parameter, and the impedance of the antenna.
		\\ \cline{2-4}	
		\hline
		\multirow{10}{1.5cm}{UAV placement and path planning} & 
		\multirow{2}{*}{\cite{m7}} & \multirow{2}{*}{PSO} & A PSO-based 3D UAV placement framework is proposed for wireless coverage of the indoor environments in high-rise buildings under disaster situations. 
		\\ \cline{2-4}&
		\multirow{2}{*}{\cite{m10}} & \multirow{2}{*}{PSO} & A PSO optimization algorithm is used to formulate the 3D UAV placement problem with QoS awareness in ad hoc wireless networks.
		\\ \cline{2-4}&
		\multirow{3}{*}{\cite{m11}} & \multirow{3}{*}{ACO, PSO} & A bio-inspired optimization learning environment is built to simulate the UAV path-planning models where SI approaches such as ACO and PSO are exploited to perform a student learning process for searching UAV paths. 
		\\ \cline{2-4}&
		\multirow{2}{*}{\cite{m12}} & \multirow{2}{*}{ACO}	& An improved ACO-based UAV path planning architecture is proposed in with obstacle taken into account in a controlled area with several radars.
		\\ \cline{2-4}	&
		\multirow{2}{*}{\cite{m13}} & \multirow{2}{*}{PSO} & A distributed PSO model is taken to generate the optimal paths for UAVs for fast convergence and minimal latency for target detection in reconnaissance contexts.
		\\ \cline{2-4}	
		\hline
		
		\multirow{8}{1.5cm}{Clustering and routing in wireless IoT networks} & 
		\multirow{2}{*}{\cite{m15}} & \multirow{2}{*}{SI} & An SI-based chicken swarm optimization framework is proposed for clustering in wireless IoT sensor networks to bring the energy efficiency improvement and energy usage balance. 
		\\ \cline{2-4}&
		\multirow{2}{*}{\cite{m16}} & \multirow{2}{*}{PSO} & A multi-objective PSO algorithm is adopted to implement the data clustering of the sanitized IoT dataset as a sensitization method to hide the confidential information.
		\\ \cline{2-4}&
		\multirow{2}{*}{\cite{m17}} & \multirow{2}{*}{PSO} & A PSO algorithm is derived in to optimize the cluster head in terms of reduction of calculation complexity and clustering decision convergence  improvement.
		\\ \cline{2-4}&
		\cite{m20} & ACO & An ACO scheme is proposed to enable a distributed intelligent traffic routing system for connected vehicles.
		\\ \cline{2-4}	&
		\multirow{2}{*}{\cite{m22}} & \multirow{2}{*}{PSO} & PSO algorithms are proposed to perform bio inspired swarm intelligence for cluster head selection with energy awareness in IoT networks. 
		\\ \cline{2-4}	
		\hline
		
		\multirow{6}{1.5cm}{Smart city applications} & 
		\cite{m24}&	PSO	&A project is proposed using PSO optimization algorithms for a smart city in the context of smart communities.
		\\ \cline{2-4}&
		\cite{m26}&	PSO	&A PSO scheme is considered for privacy guarantee in smart power systems in smart cities. 
		\\ \cline{2-4}&
		\multirow{2}{*}{\cite{m28}}	& \multirow{2}{*}{PSO} & A PSO algorithm is studied associated with support vector machine for vehicular traffic prediction in smart cities. 
		\\ \cline{2-4}&
		\multirow{2}{*}{\cite{m29}}	& \multirow{2}{*}{GWO} & A GWO technique is proposed to optimize the number of vehicular cluster for supporting vehicular clustering process. 
		\\ \cline{2-4}	
		\hline
		
		\multirow{8}{1.5cm}{Energy management in smart grid} & 
		\multirow{2}{*}{\cite{m30}} & \multirow{2}{*}{PSO} & A PSO algorithm is designed to facilitate the cost-efficient power management in multi-source renewable energy microgrids. 
		\\ \cline{2-4}&
		\multirow{2}{*}{\cite{m31}} & \multirow{2}{*}{PSO} &	A binary PSO algorithm is developed to implement the home energy management scheduling in a fashion the total power cost is minimized while satisfying user comfort. 
		\\ \cline{2-4}&
		\multirow{2}{*}{\cite{m33}}	& \multirow{2}{*}{PSO} & An evolutionary PSO-based framework is designed to solve the energy resource management problem in smart grids, by minimizing the total operational cost of energy aggregators while optimizing the power system income. 
		\\ \cline{2-4}&
		\multirow{2}{*}{\cite{m34}} & \multirow{2}{*}{GWO} & A GWO-based algorithm is developed for the planning and management of renewable energy sources with energy storage units for reducing voltage deviation and pollution caused by grid operations.
		\\ \cline{2-4}	
		\hline
	\end{tabular}
\end{table*}

Recently, an SI-based framework is designed to solve the energy resource management problem in smart grids, by minimizing the total operational cost of energy aggregators while optimizing the power system income \cite{m33}. This SI-inspired energy management model can be seen in Fig.~\ref{Fig:SI_SmartGrid}. As shown in the diagram, the energy aggregator enabled by swarm intelligence optimization  can perform efficient energy scheduling with respect to the uncertainty of energy sources such as solar energy, vehicles to grid, and electricity markets, in order to meet the various load demands. A testbed is also implemented using a PSO algorithm that proves its efficiency in implementing the flexible energy consumption management for profit maximization with multiple energy demands. GWO, another SI technique, is also applied to enable efficient energy management \cite{m34}. In this research, the planning and management of renewable energy sources with energy storage units are formulated that can be simulated by a GWO-based algorithm for reducing voltage deviation and pollution caused by grid operations. The operational planning for energy management is also studied in \cite{m35} where a joint scheme of PSO and reinforcement learning is designed to optimize the energy consumption cost of small buildings without facility considerations.  

\subsection{Summary}
\label{SubSec:Miscellaneous_Issues_Summary}
This section reviews the roles of SI in many other applications in 5G networks, including antenna design, UAV placement and path planning, clustering and routing in IoT networks, smart city applications, and energy management in smart grid. Among SI algorithms, PSO has emerged as the prominent technique for supporting effectively 5G applications in different use case scenarios. For example, the PSO can facilitate antenna design by optimizing the antenna parameters and synthesizing multi-beam patterns \cite{m1} or support microstrip antenna design that can tackle the problem of mm-wave antenna impedance mismatch in 5G communication systems \cite{m2}. Moreover, SI algorithms have proved their great potential in improving UAV-related communication designs, such as UAV placement and path planning design \cite{m7, m11, m12}. As an example, PSO can provide intelligent 3D UAV placement solutions for indoor users in small coverage areas that is promising for wireless convergence in massively crowded events.  In terms of clustering and routing solutions for IoT networks, SI has also many applications, ranging from providing IoT sensor network clustering for energy efficiency improvement and energy usage balance \cite{m15} to enabling optimized routing models for IoT ecosystems such as vehicular networks \cite{m20}. Recently, SI has been used in smart city and smart grid domains, where bio-inspired algorithms such as PSO, ACO help realize intelligent services such as vehicular traffic prediction \cite{m27} and cost-efficient power management \cite{m30}. The interesting findings from the use of SI techniques in these applications are expected to shed new lights on the future research for emerging SI-inspired 5G services. To summary this section, the reviewed literature, used techniques, and highlights are summarized in Table~\ref{Tab:Summary_Table_Miscellaneous}.
\section{{Research Challenges and Future Directions}}
\label{Sec:Challenges}
{In this section, we discuss some key research challenges and highlight possible directions on the use of SI in NGN.}

\subsection{{Research Challenges}}
 \label{SubSec:Challenges}
\subsubsection{{General Challenges Using SI Methods}}
In the preceding sections, the reason for the high applicability of SI methods is discussed. However, this does not mean that they are off-the-shelf optimizer. In other words, we need to formulate the problem in a way that is suitable for the algorithm. The majority of SI requires a designer to formulate the problem in a way to have all the parameters in a vector. This vector will be used in the algorithm to minimize or maximize an objective function. Therefore, we need to provide an objective function to evaluate each vector combination that represents a potential solution for the problem.
For instance, SI methods can be used as learning algorithms for machine learning techniques. In any learning algorithm, some parameters should be optimized to reduce the error rate. In this context, SI techniques can be employed to optimize the learning parameters. All the controlling parameters of the machine learning should be represented as a vector first \cite{mirjalili2015effective}. In neural networks, for instance, the connection weights and biases can be used in a vector and the objective function is the error rate, which is the discrepancy between the actual and expected outputs.
Another challenge when solving real-world problems using swarm intelligence methods is the existence of multiple objectives. In this case, there are multiple evaluation criteria for each solution. The objectives are typically in-conflict as well, so we have to find the best trade-offs between them. Therefore, there are no single solutions for problems with multiple objectives. A set called Pareto optimal solution is the answer for multi-objective problems \cite{mao2009evolutionary}.

\subsubsection{{Reduction in Size of Search Space}}
Since many optimization problems in NGN are defined over a large parameter search space, SI algorithms may have difficulties in finding high-quality solutions (close to global optima). Moreover, this issue becomes more challenging when the optimization problem involves both binary and continuous variables. One possible solution is to design prune policies to eliminate infeasible solutions and accelerate the optimization process \cite{huang2020bilevel}. Another solution is to initialize high-quality solutions so that SI algorithms do not fall into the trap of local points. 


\subsubsection{Ability to Find the Globally Optimum}
It is important to note that SI algorithms typically solve an optimization problem directly regardless of its convexity and optimality, i.e., they are problem-independent. After reviewing many references, we observe a common mistake that the authors usually state that their SI-based algorithms would obtain the globally optimal solutions. For instance, it is written in \cite{Jiang2020Deep} that the proposed hybrid algorithm can keep ``the advantages of the PSO in finding global optimal solutions"; however, it seems to be not correct. The fact that any SI approach maintains a balance between exploration and exploitation, and almost all SI algorithms are suitable for global optimization, but there is no guarantee for such optimality \cite{yang2014nature}. Therefore, to evaluate the global optimality of an SI-inspired algorithm, one should conduct comparisons with other global optimization techniques like branch-and-bound and exhaustive search. From the learning perspective, improving the balance between exploration and exploitation would increase the possibility of achieving the global solution. For example, the work in \cite{nguyen2019new} includes the properties of the binary search space and develops an adaptive strategy in designing a binary PSO algorithm. Moreover, recent works have investigated learning approaches to improve the exploration-exploitation balance \cite{cao2019learning}. 

There are other difficulties when solving real-world problems using swarm intelligence methods, including but not limited to, expensive objective function, discrete variables, binary variables, highly-constrained problems, noisy objective functions, etc. There are different techniques to handle each of these areas. Covering all these challenges is outside the scope of this work. In sum, although SI has great potentials and opportunities for the design and optimization and NGN, many issues should be taken into consideration when SI-inspired algorithms are investigated for practical and specific applications. 





\subsection{{Future Directions}}
\label{SubSec:ResearchDirections}
\subsubsection{{SI for Solving Large-Scale Optimization Problems}}
As the network will be very complex with massive user connection, environment dynamics, and various QoS requirements, it is really difficult to optimize such systems by SI techniques with reasonable performance and fast convergence. 
How to utilize SI techniques for the design and optimization of large-scale IoT systems is still a challenge. A promising solution, in this case, is dividing the entire network into a number of smaller regions and region heads (e.g., edge server and macro BS) are responsible for coordination. Another solution is integration with SDN and network function virtualization, which would gather and maintain global network state and information. The SI community has also developed various SI techniques for large-scale optimization. An example can be found in \cite{lan2020two} with a two-phase optimizer, which is motivated by the fact in real life that a smart leader can guide the team to have better achievements. Applying these scalable techniques would significantly improve the large-scale system performance in a comparison with conventional approaches. 

\subsubsection{{Hybrid Algorithms for Performance Improvement and Problem-solving Effectiveness}} 
The No-Free-Lunch theorem expresses that any algorithm would not find its superiority in all possible problems, that is, an algorithm is highly suitable for a problem but it may not perform well for a different one. From the reviewed literature, we observe that a hybrid algorithm typically outperforms the original algorithms if it is designed properly. For instance, it is shown in \cite{Mandloi2016aLowComplexity} that the hybrid algorithm can achieve better convergence speed and computational complexity than the original PSO and ACO methods. We should stress that a hybrid method can be designed in many ways, e.g., combining two or more basic SI techniques, adding chaotic, and controlling the balance between exploration and exploitation. Recently, many research works also consider integrating SI with deep learning techniques to accelerate and improve the solution. An example can be found in \cite{Nguyen2020Deep}, where the initial solution for a metaheuristic is approximated by a deep neural network. Simulations demonstrate that the proposed approach with deep learning can obtain almost the same performance as the original method while reducing the computational complexity up to 85\%. In sum, we expect that hybrid approaches, if designed properly, can improve the system performance significantly. 

\subsubsection{{SI for Federated Learning with Applications to Wireless and Communication Systems}}
While SI-inspired algorithms are typically deployed is a centralized fashion, many emerging applications and services should be implemented distributedly. Moreover, security and data privacy are paramount aspects that should be considered in any system. In such a case, federated learning \cite{lim2020federated} is a promising technology as the training data does not need to be transmitted to and stored at the central server, that is, the model training can be completed while the training data is kept locally at end users. Thanks to its particularities, federated learning has been recently found in many applications, especially ones at the edge networks like autonomous driving and medical services \cite{lim2020federated}. However, there are many challenges in deploying federated learning solutions such as resource allocation, communication security, and convergence guarantee for non-convex loss functions. To overcome these challenges, SI can be applied to find efficient solutions with competitive performance and convergence guarantee. 

\subsubsection{{SI for 6G Wireless Communications}}
While researches for 5G are still being undertaken, studies on 6G wireless systems have been already initialized by the research communities. A number of technologies have been proposed as key-enabling technologies in 6G, for example, THz communications, spatial modulation, reconfigurable intelligent surface, and cell-free networks \cite{dang2020should}. Moreover, many new use cases will be available, which shall include massive uRLLC and human-centric services. It is expected that AI will be the central theme for the design and optimization of 6G systems. As a branch of AI, SI shall find many applications in this network generation. For example, for a very complex problem with many constraints and optimization variables, SI can be used as a tool to generate the initial dataset, which is then used by other deep learning techniques to train and improve the system. The combination of SI with deep learning and data analytics seems to be an effective way to optimize 6G wireless systems. 

\section{Conclusion} 
\label{Sec:Conclusion}
This survey has provided the fundamentals of SI and a survey on the applications of SI techniques to NGN. Firstly, we have presented an overview of SI and several well-known SI optimizers. Secondly, we have provided a state-of-the-art review on the applications of SI for emerging issues in NGN, including spectrum management and resource allocation, wireless caching and edge computing, network security, antenna design, UAV placement and path planning, clustering and routing in IoT systems, smart city, and energy management in smart grid. Finally, we have discussed various challenges and new directions for future research. We hope that this survey will serve as a starting reference for the applications of SI to NGN, and drive great ideas in the future.


%
%



%
\balance




\end{document}